\documentclass[10.5pt,compsoc]{tst}
\usepackage{graphicx}
\usepackage{footmisc}
\usepackage{subfigure}
\usepackage{url}
\usepackage{multirow}
\usepackage[noadjust]{cite}
\usepackage{amsmath,amsthm}
\usepackage{amssymb,amsfonts}
\usepackage{booktabs}
\usepackage{color}
\usepackage{ccaption}
\usepackage{booktabs}
\usepackage{float}
\usepackage{fancyhdr}
\usepackage{caption}
\usepackage{xcolor,stfloats}
\usepackage{comment}
\graphicspath{{figures/}}
\usepackage{cuted}  
\usepackage{captionhack}
\usepackage{epstopdf}
\usepackage{graphicx}

\headevenname{\normalsize{\textbf{\emph{Tsinghua Science and Technology, xxxxx}} 20xx, 2x(x): xxx--xxx}}%
\headoddname{{\sf Jinbing Jiang et al.:}\quad {\textbf{\emph{DoF Analysis and Beamforming Design for Active IRS-aided Multi-user MIMO Wireless ...}}}}%

\newtheoremstyle{mystyle}{0pt}{0pt}{\normalfont}{1em}{\bf}{}{1em}{}
\theoremstyle{mystyle}

\newcommand{\nop}[1]{}

\makeatletter
\renewcommand{\@biblabel}[1]{[#1]\hfill}
\makeatother

\begin{document}

%
%

\hyphenpenalty=50000

\makeatletter
\newcommand\mysmall{\@setfontsize\mysmall{7}{9.5}}

\newenvironment{tablehere}
{\def\@captype{table}}
{}
\newenvironment{figurehere}
{\def\@captype{figure}}
{}

\thispagestyle{plain}%
\thispagestyle{empty}%

\let\temp\footnote
\renewcommand \footnote[1]{\temp{\normalsize #1}}
{}
\vspace*{-40pt}
\noindent{\normalsize\textbf{\scalebox{0.885}[1.0]{\makebox[5.9cm][s]
			{TSINGHUA\, SCIENCE\, AND\, TECHNOLOGY}}}}

\vskip .2mm
{\normalsize
	\textbf{
		\hspace{-5mm}
		\scalebox{1}[1.0]{\makebox[5.6cm][s]{%
				I\hfill S\hfill S\hfill N\hfill{\color{white}%
					l\hfill l\hfill}1\hfill0\hfill0\hfill7\hfill-\hfill0\hfill2\hfill1\hfill4
				\hfill \color{black}{\quad 0\hfill 1\hfill /\hfill 0\hfill 1\quad p\hfill p\hfill  1\hfill --\hfill 1\hfill 7}\hfill}}}}

\vskip .2mm
{\normalsize
	\textbf{
		\hspace{-5mm}
		\scalebox{1}[1.0]{\makebox[5.6cm][s]{%
				DOI:~\hfill~\hfill1\hfill0\hfill.\hfill2\hfill6\hfill5\hfill9\hfill9\hfill/\hfill T\hfill S\hfill T\hfill.\hfill2\hfill0\hfill 2\hfill 4\hfill.\hfill9\hfill0\hfill1\hfill0\hfill 5\hfill 6\hfill 9}}}}

\vskip .2mm\noindent
{\normalsize\textbf{\scalebox{1}[1.0]{\makebox[5.6cm][s]{%
				\color{black}{V\hfill o\hfill l\hfill u\hfill m\hfill%
					e\hspace{0.356em}xx,\hspace{0.356em}N\hfill u\hfill%
					m\hfill b\hfill e\hfill r\hspace{0.356em}x,\hspace{0.356em}%
					x\hfill x\hfill x\hfill x\hfill x\hfill%
					x\hfill x\hfill \hspace{0.356em}2\hfill0\hfill x\hfill x}}}}}\\

\begin{strip}
{\center
{\zihao{3}\textbf{
DoF Analysis and Beamforming Design for Active IRS-aided Multi-user MIMO Wireless Communication in Low-rank Channels}}
\vskip 9mm}

{\center {\sf \large
		Jinbing Jiang, Feng Shu*, Xuehui Wang, Ke Yang, Chong Shen \\Qi Zhang, Dongming Wang and Jiangzhou Wang
	}
	\vskip 5mm}
%

\centering{
	\begin{tabular}{p{160mm}}
		
		{\normalsize
			\linespread{1.6667} %
			\noindent
			\bf{Abstract:} {\sf
				Due to its ability of significantly improving data rate, intelligent reflecting surface (IRS) will be a potential crucial technique for the future generation wireless networks like 6G. In this paper, we will focus on the analysis of degree of freedom (DoF) in IRS-aided multi-user MIMO network. Firstly, the DoF upper bound of IRS-aided single-user MIMO network, i.e., the achievable maximum DoF of such a system,  is derived, and the corresponding results are extended to the case of IRS-aided multiuser MIMO by using the matrix rank inequalities. In particular, \textcolor{black}{in serious rank-deficient, also called low-rank, like line-of-sight (LoS) channel,} the network DoF may doubles over no-IRS with the help of IRS.  To verify the rate performance gain from augmented DoF, three closed-form beamforming methods, null-space projection plus maximize transmit power and maximize receive power (NSP-MTP-MRP), Schmidt orthogonalization plus minimum mean square error (SO-MMSE) and two-layer leakage plus MMSE (TLL-MMSE) are proposed to achieve the maximum DoF. Simulation results shows that IRS does make a dramatic rate enhancement. For example, in a serious rank-deficient channel, also called low-rank, the sum-rate of the proposed TLL-MMSE aided by IRS is up to 2.54 times that of no IRS. This means that IRS may make a significant DoF improvement in such a channel.}
			}
		
	\end{tabular}
}
\vskip 6mm

\vskip -3mm
\small\end{strip}

\thispagestyle{plain}%
\thispagestyle{empty}%
\makeatother
\pagestyle{tstheadings}

\begin{figure*}[b]
	\vskip -6mm
	\begin{tabular}{p{160mm}}
		\toprule\\
	\end{tabular}
	\vskip -4.5mm
	\noindent
	\setlength{\tabcolsep}{1pt}
	\begin{tabular}{p{1.5mm}p{160mm}}
		$\bullet$ & Jinbing Jiang, Ke Yang, Chong Shen and Qi Zhang are with the School of Information and Communication Engineering, Hainan University, Haikou, 570228, China. Email:JJB2511@163.com;;23220854100060@hainanu.edu.cn;chongshen@haina\\~~~~nu.edu.cn;hdzhangqi0509@163.com.\\
		$\bullet$ & F. Shu is with the School of Information and Communication Engineering, Hainan University, Haikou, 570228, China, and also with the School of Electronic and Optical Engineering, Nanjing University of Science and Technology, Nanjing, 210094, China. Email: shufeng0101@163.com.\\
		$\bullet$ & Xuehui Wang is with the School of Mathematics and Statistics, Hainan Normal University, Haikou, 571127, China. E-mail: wangxuehui0503@163.com.\\
		$\bullet$ & Dongming Wang is with the National Mobile
		Communications Research Laboratory, Southeast University, Nanjing 210096,
		China. E-mail: wangdm@seu.edu.cn.\\
		$\bullet$ & Jiangzhou Wang is with the School of Engineering, University of Kent, Canterbury CT2 7NT, U.K., Email: j.z.wang@kent.ac.uk.\\
		$\sf{*}$ & To whom correspondence should be addressed. \\
		&Manuscript received: 15-Oct-2024;
		 revised: 26-Mar-2025;
		 accepted: 22-Apr-2025
		\end{tabular}
\end{figure*}\zihao{5}

\vspace{3.5mm}
\section{Introduction}
\label{s:introduction}
\noindent
Nowadays, because of the rapid development of communication technologies, our daily lives have been greatly improved in a convenient and intelligent way. As more and more smart devices access to wireless networks, there exists a high standard requirement for rate performance and latency. However, there exist fast attention, high power-consumption, high hardware-overhead challenges for the fifth generation (5G) technologies, such as ultra-dense network, millimeter wave and multi-input multi-output (MIMO) \cite{R. Liu}-\cite{F}. Therefore, it is too difficult for 5G technologies to meet the more rigorous demand, i.e. high rate, low latency and low power consumption.

Fortunately, as the improvement of meta-materials, a novel green forwarder called passive intelligent reflecting surface (IRS) is recently proposed \cite{H. Ren}. Its surface integrates a multitude of low-cost passive reflective elements, by adjusting the phase shift of each element, IRS can intelligently control the wavefront of the incident signals in a desired manner. In terms of hardware cost, rate gain and channel reconfigurability, IRS has many advantages such as low circuit cost,  high energy efficiency, rate enhancement, and coverage extension. Thus, it is viewed as one of the most potential key techniques for the future wireless networks like 6G, UAV communications, satellite communications, and marine communications. Because of the unique features, IRS can be applied to wireless networks for coverage extension and energy conservation \cite{C. Huang}-\cite{P. Wang}. Besides that, the hot noise introduced in IRS can be negligible \cite{M. Di}. For these reasons, more and more researchers pay attention to IRS \cite{SH. Zhang}-\cite{E. Basar}. Many authors used the advantage of IRS to control the direction of incident signals to assist communication networks for secure transmission \cite{L. Yang}-\cite{L. Dong}. In an IRS-assisted directional modulation (DM) network, the authors in \cite{Y. Lin} designed the beamforming to enhance the secrecy rate performance.  To enhance the energy harvesting performance (EHP), an IRS-aided simultaneous wireless information and power transfer (SWIPT) wireless network was put forward in \cite{C. Pan}, while it was demonstrated that the EHP of the SWIPT system could be enhanced by IRS. In \cite{T. Bai}, IRS was adopted to mobile edge computing  system to reduce the computational latency. The authors employed IRS to cognitive radio systems in \cite{L. Zhang}, where the problem of maximizing the weighted sum-rate (WSR) of secondary users (SUs) was formulated. By jointly optimizing SU transmit precoding and IRS phase shifts, it was showed that the significant WSR improvement could be attained with the aid of IRS. Moreover, by controlling the incident light beam, IRS could extend the range of rotation angle of the visible light communication receiver \cite{A. R. Ndjiongue}, while the field-of-view was improved. An IRS-aided decode-and-forward (DF) relay network was designed in \cite{X. Wang}, where a alternately iterative structure was proposed to obtain the analysis solutions to DF beamforming vector and IRS reflection matrix, so that the receive power was maximum.

So far, passive IRS have been well applied in different communication scenarios, such as secure communication, high-precision positioning and electromagnetic pollution. However, passive IRS faces a path loss (PL) challenge in fact \cite{M. H. Khoshafa}. Usually, passive IRS is intelligently configured between base station (BS) and user to improve communication quality. The received signals reflected by IRS experience two links namely BS-IRS link and IRS-user link, the PL of cascaded link (PLCL) is the product of the PLs corresponding to these two links, rather than the sum of them. Generally, the PLCL is thousands of times larger than the PL of the direct link \cite{M. Najafi}. With further research on passive IRS, an innovative technology called active IRS is proposed. The key distinction between passive IRS and active IRS is that passive IRS only passively reflect the incident signal without amplification, while each active IRS element integrates a reflective amplifier, which makes that active IRS not only reflects the incident signal but also amplifies it with additional power \cite{Z. Zhang}. Therefore, active IRS can eliminate the fundamental physical limitation namely PLCL \cite{L. Cantos}-\cite{M. Fu}.

Based on these unique advantages, a lot of research on active IRS has been implemented \cite{Y. Wang}-\cite{S. Gong}. The authors in \cite{K. Zhi} compared the active and passive IRS-aided system theoretically and numerically, and the simulation results showed that active IRS was better than passive when the power budget was not so small and the number of IRS elements was not so large. In \cite{A. R.}, the authors have demonstrated that introducing active IRS to wireless network can obtain rate improvement compared to passive IRS-aided network. An active IRS-aided multi-user multiple-input multiple-output (MU-MIMO) was presented in \cite{Y. Ma}, where energy efficiency was maximized by jointly designing transmit beamforming and energy-included IRS reflection matrix. The authors in \cite{X. Yang} investigated a new active IRS-aided uplink multi-antenna non-orthgonal multiple access system, base on minimum mean square error (MMSE) criterion, the proposed system achieved higher sum-rate than those of various benchmarks by an efficient alternating optimization (AO) algorithm. With the practical hardware impairments, \cite{Z. Peng} considered an active IRS-assisted the multiuser downlink transmission systems. By designing BS transmit beamforming and active IRS reflection matrix including the amplification factors and the phase shifts, the maximum sum date rate could be achieved. To improve the rate of a wireless network aided by an active IRS, \cite{Y. Line} first simplified signal-to-noise ratio (SNR), and applied Rayleigh-Ritz theorem to maximize SNR by designing active IRS reflection matrix. To improve rate, generalized maximum ratio reflection method and maximizing SNR method based on fractional programming were respectively proposed. \textcolor{black}{The authors in \cite{K. Yang} proposed a  multi-IRS-aided multi-stream DM network. To eliminate the interference among different data streams, the null-space projection (NSP) at BS and receive zero-forcing (ZF) at user is adopted to design the beamforming of transmitter and receiver, and phase alignment (PA) are adopted to obtain the phase shift matrix (PSM). In \cite{D. Xu}, an active IRS-assisted multi-user system was considered, and a method of minimizing the BS transmit power was proposed to jointly optimizes the reflection matrix at IRS and the beamforming vectors at BS to achieve a green wireless communication. \cite{M. Munawar} proposed a low-complexity adaptive selection beamforming scheme for the practical implementation of IRS-assisted single-user wireless networks. Compared with AO, the proposed scheme requires lower computational cost.}

However, it is difficult to obtain the precise channel capacity of IRS-assisted MU-MIMO communication networks. DoF is a performance metric, which is easier to be analyzed and handled. In the case of high SNR, degree of freedom (DoF) main shows how the channel capacity scales \cite{H. V. Cheng}. For IRS-aided MU communication network, DoF has been well investigated by many researchers \cite{K. G. Seddik}. 
In \cite{H. Do}, it was proved that as IRS aperture increased, DoF increased, which did not need the growth of transmitter, receiver and multipath propagation. A wiretap channel of an IRS-aided secure MIMO wireless network was considered in \cite{M. Nafea}, where the lower and upper bounds of the secure DoF were respectively derived. Compared to a separate external multi-antenna cooperative jammer, the secure rate could be improved by designing IRS reflection matrix. For active IRS-aided time-selective $K$-user interference channel (IC), the DoF region and the sum DoF were derived in \cite{A. H. A. Bafghi}, respectively. Meanwhile, it was proved that when the scale of active IRS elements was large, the sum DoF is $K$. Then, the results of active IRS was extended to the case of passive IRS. Moreover, it was demonstrated that the lower bound of sum DoF was close to $K$ if IRS elements was large-scale. For an arbitrary antenna configuration of transmitter and receiver, \cite{S. Zheng} verified that active IRS could enhance the sum DoF of a two-user MIMO IC. Via combining IRS beamforming, zero-forcing transmission and interference decoding, the interference problem was solved. Subsequently, the authors derived the maximum sum DoF and the requirement for obtaining IRS gain.

However, $K$-user IC in \cite{A. H. A. Bafghi} is time-selective, and BS and users  are both single-antenna. Furthermore, the users in \cite{S. Zheng} are only two, their practical applications are limit. How about the achievable maximum DoF of a general IRS-aided wireless network? In this paper, we will  make an investigation  on the achieved maximum number of DoFs of a general active IRS-aided multi-user MIMO wireless network with multiple transmit antennas at BS and receive multiple antennas per user.  The main contributions of this paper are summarized as follows:

\begin{enumerate}
	\item  To evaluate the DoF gains achieved by active IRS, an IRS-aided multi-user network is established and the corresponding achievable DoF upper bounds in single-user MIMO network \textcolor{black}{is} derived in \textcolor{black}{rank-deficient} channels. Then, the corresponding results are also extended to the multi-user scenarios. In particular, it is found that introducing IRS will make a dramatic DoF enhancement. For example, in \textcolor{black}{a general} channel, the number of DoFs of IRS-aided wireless network is twice that of no IRS.  However, for LoS channel, \textcolor{black}{as a low-rank channel, the} corresponding DoF gain will become trivial as the number $K$ of users tends to medium or large-scale. When the $K$ goes to 1, the DoF enhancement is close or equal to that of the \textcolor{black}{rank-deficient} channel. Finally, the special condition for achieving the maximum DoF of single-user in LoS channel is constructed, i.e. channels orthogonal condition (OC) to achieve two DoFs .
 
	\item  To confirm the sum-rate gains achieved by incremental DoF created by IRS in \textcolor{black}{rank-deficient} channel, a two-layer leakage plus minimum mean square error (TLL-MMSE) method is proposed, where the precoding vectors and receive beamforming vectors are designed by the leakage concept and MMSE rule, respectively, while the phase is adjusted by the leakage principle.  Simulation results shows that the sum-rate of the proposed TLL-MMSE with IRS may make a significant sum-rate improvement over no IRS, i. e., achieving up to \textcolor{black}{2.54 times} gain over no IRS.
	
	\item Finally, to verify the DoF gains achieved by IRS  in LoS channel, two beamforming methods, called null-space projection plus maximize transmit power and maximize receive power (NSP-MTP-MRP) and schmidt orthogonalization plus the minimum mean squared error (SO-MMSE), are proposed. For both methods, phase alignment is adopted to implement the phase adjustment. Simulation results indicate that the sum-rates of  the proposed SO-MMSE  and NSP-MTP-MRP with IRS  still may achieve up to more than 140\% rate gain over those of no IRS. Moreover,  the former performs much better than the latter in terms of sum-rate.
\end{enumerate}

The remaining content is  as follows. In section 2 the system model is shown. The DoF analysis and condition for achieving the DoF upper in LoS channel are shown in section 3. Section 4 denotes the proposed three methods. Simulation results we obtained are presented in section 5. Finally, conclusions are provided in section 6. 

Notations: In this paper, boldface lower case and upper case letters represent vectors and matrices, respectively. The sign $\mathbb{C} $ presents set of complex. The notations $\left | \cdot  \right | $, $\| \cdot \|$and$\| \cdot \|_F$ stand for the determinant, 2-norm and F-norm operations, respectively. Signs $[\cdot]^*$, $[\cdot]^{\dagger}$, $[\cdot]^T$, $[\cdot]^{-1}$ and $[\cdot]^{H}$ express the conjugate, pseudo-inverse, transpose, inverse and conjugate-transpose operations, respectively. the diagonal operator is presented as $\text{diag}(\cdot)$. The $\text{tr}(\cdot)$ denote the trace operation. The sign $\mathbb{E} [\cdot]$ express expectation. The sign $\mathcal {CN}(\mu,\sigma ^2)$ signifies the complex Gaussian distribution with mean $\mu$ and variance $\sigma ^2$.

\section{System Model}
\label{s:System Model}
\begin{figure}
	\centering
	\includegraphics[width=1\linewidth]{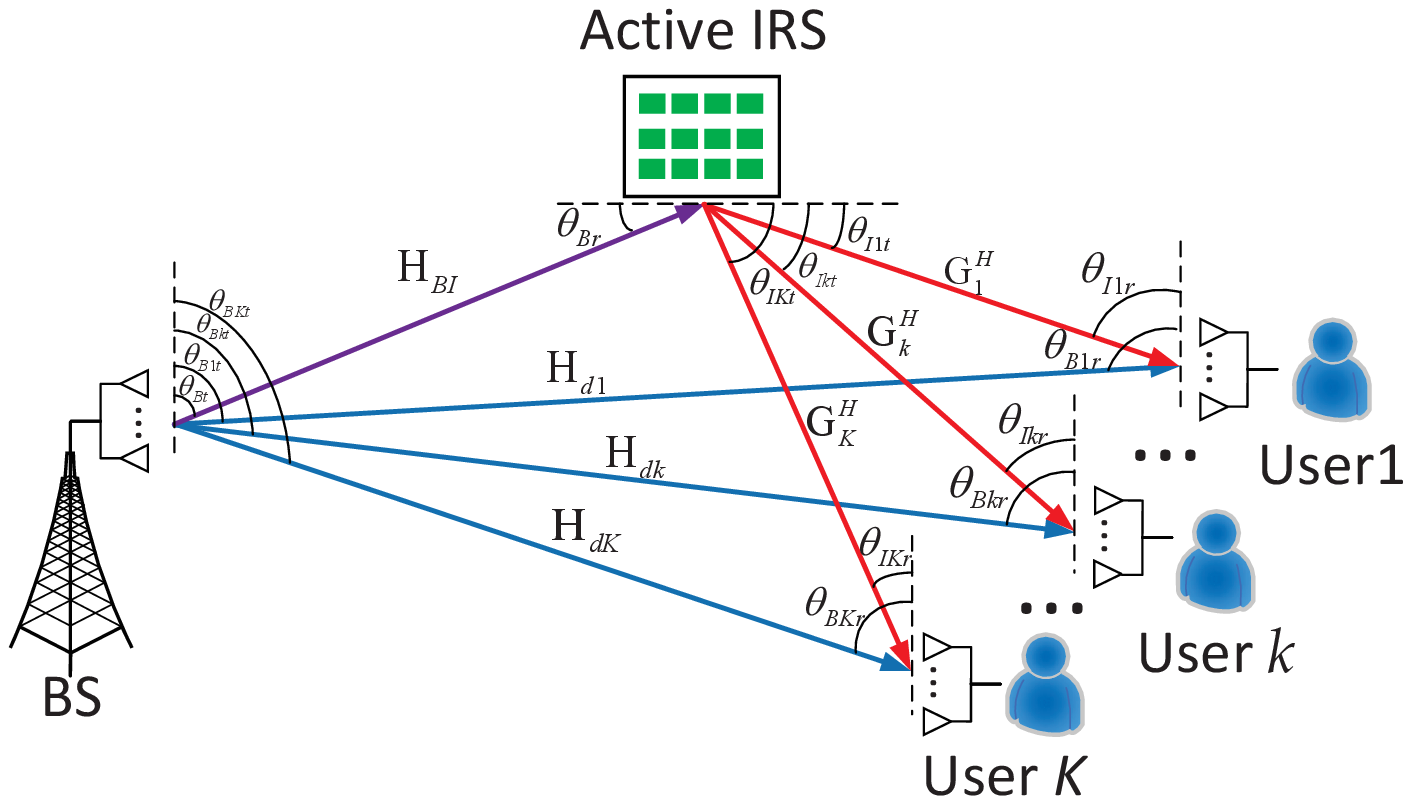}
	\caption{System model of active IRS-aided multi-user MIMO network}
	\label{fig:DoFLoS}
\end{figure}

\noindent
In Fig.~\ref{fig:DoFLoS}, an active IRS-aided-multi-user wireless communication network is plotted with $M$ transmit antennas, $K$ users and an IRS, where the user $k$ and IRS are equipped $Q_k$ antennas and $N$ reflective elements respectively, where $k=1,...,K$. Let us define $N_U =  \sum_{k= 1}^{K}Q_k$, and $N_U$ is the sum number of antennas of all users.
The transmit baseband signal is written by
\begin{align}
	\mathbf{x}=\sum_{k=1}^{K}\mathbf{V}_k\mathbf{s}_k,
\end{align}	
where $\mathbf{V}_k \in \mathbb{C}^{M \times L_k}$ is the $k$-th transmission beamforming matrix at BS, and $\mathbf{s}_k\in \mathbb{C}^{L_k \times 1}$ is the corresponding data symbol, where $\mathbb{E}\{\mathbf	{s}_k^H\mathbf{s}_k\}=\mathbf{I}_{L_k}$. The reflect signal of the IRS is
\begin{align}
	\mathbf{y}^t=\mathbf{\Theta}\mathbf{H}_{BI}\mathbf{x}+\mathbf{\Theta}\mathbf{n},
\end{align}	
where $\mathbf{H} _{BI}\in \mathbb{C}^{N \times M}$ denotes the  channel matrix from BS to IRS, and $\mathbf{\Theta}=\mathrm{diag}(\alpha _1e^{j\theta _1},...,\alpha _{N}e^{j\theta _{N}})\in \mathbb{C}^{N \times N}$ stands for the reflection coefficient matrix of IRS, where $\alpha _l$ represents amplitude gain and $\theta _l$ stands for phase shift, respectively. And $\mathbf{n} \in \mathbb{C}^{N \times 1}$ is the additive white Gaussian noise (AWGN) at IRS, and $\mathbf{n} \sim  \mathcal {CN}(0,\sigma^2_{irs} \mathbf{I}_{N}) $.

The receive signal at user $k$ is represented as
\begin{align}\label{3}
	\mathbf{y}_k^r =& \mathbf{H}_{dk}\sum_{i=1}^{K}\mathbf{V}_i\mathbf{s}_i+\mathbf{G}_k^H\mathbf{\Theta}\mathbf{H}_{BI}\sum_{i=1}^{K}\mathbf{V}_i\mathbf{s}_i\nonumber\\&+\mathbf{G}_k^H\mathbf{\Theta}\mathbf{n}+\mathbf{z}_k,
\end{align}
where $\mathbf{G}_k^H  \in \mathbb{C}^{Q_k \times N}$ and $\mathbf{H}_{dk} \in \mathbb{C}^{Q_k \times M}$ denote the channel gain matrices from IRS to user $k$ and  BS to user $k$, respectively, $\mathbf{z}_k \in \mathbb{C}^{Q_k \times 1}$ is the additive white Gaussian noise (AWGN) at user $k$, and $\mathbf{z}_k \sim  \mathcal {CN}(0,\sigma_z ^2 \mathbf{I}_{N})$. After the receive beamforming is performed, we have
\begin{align}
	\mathbf{y}_k &= \mathbf{U}_k^H\mathbf{H}_{dk}\mathbf{V}_k\mathbf{s}_k+\mathbf{U}_k^H \mathbf{G}_k^H\mathbf{\Theta}\mathbf{H}_{BI}\mathbf{V}_k\mathbf{s}_k\nonumber\\&+\mathbf{U}_k^H\mathbf{H}_{dk}\sum_{i=1,i\neq k}^{K}\mathbf{V}_i\mathbf{s}_i+\mathbf{U}_k^H\mathbf{G}_k^H\mathbf{\Theta}\mathbf{n}\nonumber\\&+\mathbf{U}_k^H\mathbf{z}_k+\mathbf{U}_k^H\mathbf{G}_k^H\mathbf{\Theta}\mathbf{H}_{BI}\sum_{i=1,i\neq k}^{K}\mathbf{V}_i\mathbf{s}_i,
\end{align}
where $\mathbf{U}_k^H \in \mathbb{C}^{L_k \times Q_k}$ is the receive beamforming matrix , .

Then the average receive power  at user $k$ is defined by
\begin{align}
	&\mathbb{E}\{\mathbf	{y}_k^H\mathbf{	y}_k \}= \|\mathbf{U}_k^H\mathbf{H}_{dk}\mathbf{V}_k\mathbf{V}_k^H\mathbf{H}_{dk}^H\mathbf{U}_k\|_F\nonumber\\
	&+\|\mathbf{U}_k^H\mathbf{G}_k^H\mathbf{\Theta}\mathbf{H}_{BI}\mathbf{V}_k\mathbf{V}_k^H\mathbf{H}_{BI}^H\mathbf{\Theta}^H\mathbf{G}_k\mathbf{U}_k\|_F\nonumber\\
	&+\sum_{i=1,i\neq k}^{K}\|\mathbf{U}_k^H\mathbf{H}_{dk}\mathbf{V}_i\mathbf{V}_i^H\mathbf{H}_{dk}^H\mathbf{U}_k\|_F\nonumber\\
	&+\sum_{i=1,i\neq k}^{K}\|\mathbf{U}_k^H\mathbf{G}_k^H\mathbf{\Theta}\mathbf{H}_{BI}\mathbf{V}_i\mathbf{V}_i^H\mathbf{H}_{BI}^H\mathbf{\Theta}^H\mathbf{G}_k\mathbf{U}_k\|_F\nonumber\\
	&+\sigma_{irs} ^2\|\mathbf{U}_k^H\mathbf{G}_k^H\mathbf{\Theta} \mathbf{\Theta}^H\mathbf{G_k}\mathbf{U}_k\|_F+\sigma ^2_z\|\mathbf{U}_k^H\mathbf{U}_k\|_F.
	\end{align}

{Then the corresponding rate can be obtained as} 
\begin{align}
	\mathbf{\gamma}_k&=\log_2\left|\mathbf{I}_k+(\mathbf{A}_1+\mathbf{A}_2)(\mathbf{B}_1+\mathbf{B}_2+\mathbf{B}_3)^{-1}\right| ,
\end{align}
where 

\begin{align}
	\mathbf{A}_1&=\mathbf{U}_k^H\mathbf{H}_{dk}\mathbf{C}\mathbf{H}_{dk}^H\mathbf{U}_k,\nonumber\\ \mathbf{A}_2&=\mathbf{U}_k^H\mathbf{G}_k^H{\mathbf{\Theta}}\mathbf{H}_{BI}\mathbf{C}\mathbf{H}_{BI}^H{\mathbf{\Theta}}^H\mathbf{G}_k\mathbf{U}_k,\nonumber\\
	\mathbf{B}_1&=\sum_{i=1,i\neq k}^{K}\mathbf{U}_k^H\mathbf{H}_{dk}\mathbf{C}\mathbf{H}_{dk}^H\mathbf{U}_k,\nonumber\\
	\mathbf{B}_2&=\sum_{i=1,i\neq k}^{K}\mathbf{U}_k^H\mathbf{G}_k^H{\mathbf{\Theta}}\mathbf{H}_{BI}\mathbf{C}\mathbf{H}_{BI}^H{\mathbf{\Theta}}^H\mathbf{G}_k\mathbf{U}_k,\nonumber\\
	\mathbf{B}_3&={\sigma^2_{irs}}  \mathbf{U}_k^H\mathbf{G}_k^H{\mathbf{\Theta}} {\mathbf{\Theta}}^H\mathbf{G_k}\mathbf{U}_k+\sigma ^2_z\mathbf{U}_k^H\mathbf{U}_k,\nonumber\\
	\mathbf{C}&=\mathbf{V}_k\mathbf{V}_k^H.
\end{align}

So the sum-rate of all users is defined as follows
\begin{align}
	\mathbf{\gamma}_{sum}&=\sum_{k=1}^{K}\mathbf{\gamma}_k.
\end{align}
\section{DoF Analysis}
\noindent
In this section, we will present a proof of the achievable upper bounds of DoF  in SU-MIMO and MU-MIMO networks with the aid of IRS by using matrix rank inequalities. 

\subsection{DoF analysis}
\noindent
In Fig.~\ref{fig:DoFLoS}, based on (\ref{3}),  the signal at user $k$ is represented as
\begin{align}
	\mathbf{y}_k=( \mathbf{\underbrace{\mathbf{G}_k^H \mathbf{\Theta} \mathbf{H}_{BI}}_{\mathbf{A}}}+\mathbf{H}_{dk})\mathbf{x}+ \mathbf{G}_k^H \mathbf{\Theta} \mathbf{n}+{\mathbf{z}_k}.
\end{align}

Let us define 
\begin{align}\label{TotalChannel}
	\mathbf{\hat{H}} & =	\mathbf{\underbrace{\mathbf{G}_k^H \mathbf{\Theta} \mathbf{H}_{BI}}_{\mathbf{A}}}+\mathbf{H}_{dk}.
\end{align}
In accordance with the matrix rank inequality \cite{Horn}
\begin{align}\label{rank-A+B-C}
	\mathrm{rank}\left(\mathbf{A}+\mathbf{B}\right)\leq \mathrm{rank}\left(\mathbf{A}\right)+\mathrm{rank}\left(\mathbf{B}\right),
\end{align}
with equality in the second inequality if and only if $(\mathrm{range}  \left(\mathbf{A}\right))\cap(\mathrm{range} \left(\mathbf {B}\right))=\{0\}$ and { $(\mathrm{range} \left(\mathbf{A}^T\right))\cap(\mathrm{range}\left(\mathbf{B}^T\right))=\{0\}$}.We have
\begin{align}
	&\mathrm{rank}\left(\mathbf{H}_{dk}+\mathbf{G}_k^H \mathbf{\Theta}_k \mathbf{H}_{BI}\right)\nonumber\\
	\leq &\mathrm{rank}\left(\mathbf{G}_k^H \mathbf{\Theta} \mathbf{H}_{BI}\right)+\mathrm{rank}\left(\mathbf{H}_{dk}\right).
\end{align}

In accordance with 
\begin{align}\label{rank-AB}
	\mathrm{rank}\left(\mathbf{AB}\right)\leq \mathrm{min}\left(\mathrm{rank} \left(\mathbf{A}\right), \mathrm{rank}\left( \mathbf{B}\right)\right),
\end{align}
we have
{\begin{align}
	&\mathrm{rank}\left(\mathbf{G}_k^H \mathbf{\Theta} \mathbf{H}_{BI}\right)\nonumber\\
	&\leq \mathrm{min}\left(\mathrm{rank} (\mathbf{G}_k^H\mathbf{\Theta}), \mathrm{rank} \left(\mathbf{H}_{BI}\right)\right)\nonumber\\
	&\leq	\mathrm{min} \left(\mathrm{rank} (\mathbf{G}_k^H),\mathrm{rank}\left( \mathbf{\Theta}\right), \mathrm{rank} \left(\mathbf{H}_{BI}\right)\right).
\end{align}}

Considering the matrix $\mathbf{\Theta}$ is a full-rank diangonal matrix, the above inequality reduces to
\begin{align}\label{rank-ABC}
	\mathrm{rank}\left(\mathbf{G}_k^H \mathbf{\Theta} \mathbf{H}_{BI}\right)\leq \mathrm{min}(\mathrm{rank} (\mathbf{G}_k^H), \mathrm{rank} (\mathbf{H}_{BI})).
\end{align}

Substituting  the above expression in yields 
\begin{align}
	&\mathrm{rank}\left(\mathbf{H}_{dk}+\mathbf{G}_k^H \mathbf{\Theta} \mathbf{H}_{BI}\right)\nonumber\\
	&\leq \mathrm{min}\left(\mathrm{rank} (\mathbf{G}_k^H), \mathrm{rank} (\mathbf{H}_{BI})\right)+\mathrm{rank}\left(\mathbf{H}_{dk}\right),
\end{align}
where  the equality holds when 
\begin{align}
	\mathbf{H}_{dk}\mathbf{G}_k^H=\mathbf{0},
\end{align}
in accordance with the equality condition in (\ref{rank-A+B-C}).

In LoS channel, $\mathrm{rank}\left(\mathbf{G}_k^H\right)=\mathrm{rank}\left(\mathbf{H}_{dk}\right)=\mathrm{rank}\left(\mathbf{H}_{BI}\right)=1$, so we have the following inequality
\begin{align}
	\mathrm{rank}\left(\mathbf{H}_{dk}+\mathbf{G}_k^H \mathbf{\Theta} \mathbf{H}_{BI}\right)\leq 2,
\end{align}
whose equality condition will be specified in the next subsection.

\textcolor{black}{In full-rank channel},
\begin{align}
	\mathrm{rank}\left(\mathbf{G}_k^H\right)&=\mathrm{min}(N, Q_k),\nonumber\\
	\mathrm{rank}\left(\mathbf{H}_{dk}\right)&=\mathrm{min}(M,Q_k),\nonumber\\ \mathrm{rank}\left(\mathbf{H}_{BI}\right)&=\mathrm{min}(N,M),
\end{align}
where all elements of matrices $\mathbf{G}_k^H$, $\mathbf{H}_{dk}$ and $\mathbf{H}_{BI}$ are independent and identically distributed (i.i.d.), so we have the following inequality
\begin{align}
	\mathrm{rank}\left(\mathbf{H}_{dk}+\mathbf{G}_k^H \mathbf{\textcolor{black}\Theta} \mathbf{H}_{BI}\right)\leq \mathrm{min}(M, Q_k) \nonumber\\ +\mathrm{min}\left(\mathrm{min}(N, Q_k), \mathrm{min}(N, 
	M)\right).
\end{align}

Now, we turn to the multi-user scenario. Supposing there are $K$ users with  user
$k$ having $Q_k$ antennas. Channel matrices $\mathbf{G}_k^H$ and $\mathbf{H}_{dk}$ stand for the channel gain matrices from IRS to user $k$ and BS to user $k$.
Based on (\ref{3}), the total signal received at user $k$ is 
\begin{align}
	\mathbf{y}_k=&\left( \sum_{k= 1}^{K}\mathbf{H}_{dk}+\left(\sum_{k= 1}^{K}\mathbf{G}_k^H \mathbf{\textcolor{black}\Theta}\right) \mathbf{H}_{BI}\right)\mathbf{x} \nonumber\\+&\sum_{k= 1}^{K} \mathbf{G}_k^H \mathbf{\textcolor{black}\Theta} \mathbf{n}+\mathbf{z}.
\end{align}

In accordance with the (\ref{rank-AB}) and (\ref{rank-ABC})
\begin{align}\label{rank-m}
	&\mathrm{rank}\left(\sum_{k= 1}^{K}\mathbf{G}_k^H \mathbf{\textcolor{black}\Theta} \mathbf{H}_{BI}\right) \nonumber \\
	&\leq \mathrm{min}(\mathrm{rank}( \sum_{k= 1}^{K}\mathbf{G}_k^H),\mathrm{rank}\left(\mathbf{ H}_{BI}\right)),
\end{align}
where $k=1,\dots,K$.

Substituting  the above expression in yields
\begin{align}
	\mathrm{rank}\left(\sum_{k= 1}^{K} \mathbf{H}_{dk}\right)\le \sum_{k= 1}^{K}\mathrm{rank}( \mathbf{H}_{dk}),
\end{align}
so, it can be further obtained 
\begin{align}\label{DOF}
	&\mathrm{rank}\left(\sum_{k= 1}^{K}\mathbf{H}_{dk}+\left(\sum_{k= 1}^{K}\mathbf{G}_k^H \mathbf{\textcolor{black}\Theta}\right) \mathbf{H}_{BI}\right)\\
	&\le \sum_{k= 1}^{K}\mathrm{rank}( \mathbf{H}_{dk})+\mathrm{min}(\mathrm{rank}( \sum_{k= 1}^{K}\mathbf{G}_k^H),\mathrm{rank}(\mathbf{ H}_{BI})).\nonumber
\end{align}

In LoS channel, we have\textcolor{black}{
\begin{align}
 &\mathrm{rank}\left(\mathbf{H}_{d1}\right)=\dots=\mathrm{rank}\left(\mathbf{H}_{dK}\right)=1,
\end{align}}
so we have the following inequality
\begin{align}
	\mathrm{rank}\left(\sum_{k= 1}^{K}\mathbf{H}_{dk}+\left(\sum_{k= 1}^{K}\mathbf{G}_k^H \mathbf{\textcolor{black}\Theta} \right)\mathbf{H}_{BI}\right)	\le K+1.
\end{align}

In what follows, we consider another special scenario, called \textcolor{black}{rank-deficient} channel, \textcolor{black}{defined as} 
\begin{align}
	\mathrm{rank}\left(\mathbf{H}_{dk}\right)&=I_k\le  Q_k,
\end{align}
	\textcolor{black}{with}
	\begin{align}
	\mathrm{rank}\left(\mathbf{G}_k^H\right)&=J_k\le  Q_k,\nonumber\\
	 \mathrm{rank}\left(\mathbf{H}_{BI}\right)&=\mathrm{min}(N,M),
\end{align}
where $\mathbf{G}_k^H$ \textcolor{black}{and} $\mathbf{H}_{dk}$ are rank-deficient and independent with each other with $Q_k \ge I_k+J_k$, due to \textcolor{black}{the fact that there are  a few reflecting or scattering clusters, i.e., no rich-scattered environment}, so they have the following inequality
\begin{align}
	&\mathrm{rank}\left(\sum_{k=1}^K \mathbf{H}_k\right)\nonumber\\
	\le &\sum_{k= 1}^{K}I_k+\mathrm{min}\left(\mathrm{rank}\left( \sum_{k= 1}^{K}J_k,N,M\right)\right),
\end{align}
which gives the achievable upper bound of IRS-aided MU-MIMO systems, \textcolor{black}{where $\mathbf{H}_{k}=\mathbf{H}_{dk}+\mathbf{G}_k^H \mathbf{\Theta}\mathbf{H}_{BI}$}. In other words, the bound will determine the maximum number of bit streams parallelly transmitted from BS to all users with the aid of IRS. 

Considering that $\mathbf{H}_k$ is a $Q_k\times M$ channel matrix, so 
\begin{align}
	&\mathrm{rank}\left(\mathbf{H}_{k}\right)
	\le \mathrm{min}\left(Q_k,M\right),
\end{align}
where $M \gg Q_k$, i.e. $\mathrm{rank}\left(\mathbf{H}_{k}\right)
\le \mathrm{min}\left(Q_k\right)$.

When the sum number of antennas of all user is smaller than $N$ and $M$, and $Q_k \ge I_k+J_k$, we have a simple DoF inequality as follows
\begin{align} \label{DoF}
	\mathrm{rank}\left(\sum_{k=1}^K \textcolor{black}{\mathbf{H}_k}\right) \leq \sum_{k=1}^K I_k + \sum_{k=1}^K J_k,
\end{align}
\textcolor{black}{with equality if and only if $\mathrm{range}(\mathbf{H}_{dk})\cap \mathrm{range}(\mathbf{G}_{k}^H\mathbf{\Theta}\mathbf{H}_{BI})=\{0\}$ and $\mathrm{range}(\mathbf{H}_{dk}^H)\cap \mathrm{range}(\mathbf{H}_{BI}^H\mathbf{\Theta}^H\mathbf{G}_{k})=\{0\}$}.

When $I_k=J_k$, the above inequality reduces to
\begin{align}
	&\mathrm{rank}\left(\sum_{k=1}^K \textcolor{black}{\mathbf{H}_k}\right)	\le 2\sum_{k= 1}^{K}I_k.
\end{align}
 Without the aid of IRS, we have
\begin{align}\label{rl}
	\mathrm{rank}\left(\sum_{k= 1}^{K} \mathbf{H}_{dk} \right)	\le \sum_{k= 1}^{K}I_k.
\end{align}

Observing the above two results, it very clear that \textcolor{black}{compared with no-IRS MU-MIMO network, the upper bound of DoF of the IRS-aided MU-MIMO network doubles in a rank-deficient channel due to the help of IRS.}

\textcolor{black}{Specially, $I_k=1$ and $J_k=Q_k$, i.e., $\mathbf{H}_{dk}$ and $ \sum_{k= 1}^{K} \mathbf{G}_k^H \mathbf{\Theta}_k \mathbf{H}_{BI}$ are low-rank and full-rank channels, respectively. Then (\ref*{DoF}) can be written as
\begin{align}\label{with IRS1}
	\mathrm{rank}\left(\sum_{k= 1}^{K} \mathbf{H}_{k}\right)	\le \sum_{k= 1}^{K}Q_k.
\end{align}}

\textcolor{black}{Without the aid of IRS, we have
\begin{align}\label{no IRS1}
	\mathrm{rank}\left(\sum_{k= 1}^{K} \mathbf{H}_{dk} \right)	\le K,
\end{align}
it is known that the upper bound of DoF of the IRS-aided MU-MIMO network is $(\sum_{k= 1}^{K}Q_k)/K$ times that of no IRS a rank-deficient channel due to the help of IRS.}


\subsection{Condition for achieving the maximum DoF in LoS channel}
\noindent
In accordance with  (\ref{rank-A+B-C}), when K=1,  the maximum DoF of network as shown in Fig. 1 is two  with the orthogonal condition  (OC) 
\begin{align}\label{OC}
	\mathbf{H} _{BI}\mathbf{H} _{dk}^H=0.
\end{align}

In the following, we will show what is its specific form in a LoS channel.
As we know, in a LoS channel, the corresponding channel gain matrices in Fig.~1 have the following  rank-one form 
\begin{align}\label{8}
	\mathbf{H} _{BI}&=\mathbf{h}(\theta _{Br})\mathbf{h}^H(\theta _{Bt})\in \mathbb{C}^{N \times M},\nonumber\\
	\mathbf{H} _{dk}&=\mathbf{h}(\theta _{Bkr})\mathbf{h}^H(\theta _{Bkt})\in \mathbb{C}^{Q_k \times M},\nonumber\\
	\mathbf{G} _{k}^H&=\mathbf{h}(\theta _{Ikr})\mathbf{h}^H(\theta _{Ikt})\in \mathbb{C}^{Q_k \times N},
\end{align}
where the normalized steering vector $\mathbf {h}(\theta)$ is defined as
\begin{align}\label{h}
	\mathbf{h}(\theta)=\frac{1}{\sqrt{N} }\left [ e^{j2\pi \Phi _\theta(1)},...,e^{j2\pi \Phi _\theta(n)},...,e^{j2\pi \Phi _\theta(N)} \right ]^H, 
\end{align}
where the phase shift $\Phi_n(\theta)$ is defined as
\begin{align}
	\Phi _N(\theta)=-\frac{d}{\lambda } (n-1)\cos\theta,~~~~~~n=1,...,N,
\end{align}
where $\lambda $, $n$, $d$ and $\theta$ denote the wavelength, the antenna
index, the element spacing in the transmit antenna array
and the direction of departure, respectively. 
According to (\ref 8) , we have 
\textcolor{black}{
\begin{align}
	\mathbf{H} _{BI}\mathbf{H} _{dk}^H&=\mathbf{h}(\theta _{Br})\mathbf{h}^H(\theta _{Bt})\mathbf{h}(\theta _{Bkt})\mathbf{h}^H(\theta _{Bkr})\nonumber\\
	&=\left(\mathbf{h}^H(\theta _{Bt})\mathbf{h}(\theta _{Bkt})\right)\left(\mathbf{h}(\theta _{Br})\mathbf{h}^H(\theta _{Bkr})\right).
	\end{align}} 

Substituting the equation in  (\ref {OC}) , we have the following simple OC 
\begin{align}
	\mathbf{h}^H(\theta _{Bt})\mathbf{h}(\theta _{Bkt})=0, 
\end{align}
which means
\begin{align}
	\cos(\theta_{Bt})-\cos(\theta_{Bkt}) & = \frac{\lambda i}{Md },
\end{align} 
in accordance with the conclusions in  \cite{Tse}. The above expression is rewritten as
\begin{align}
	\theta_{Bt}=\arccos(\cos(\theta_{Bkt})+\frac{\lambda i}{Md }).
\end{align}

When $\theta_{Bkt}$ is set to $60^0$,  we have
\begin{align}
	\cos(\theta_{Bt} )= \frac{\lambda i}{Md }+\frac{1}{2 }.
\end{align}

Let us set $i=1$, $d=\frac{\lambda}{2}$ and $M=16$, which directly gives  
\begin{align}\label{Bt}
	\theta_{Bt}\approx \textcolor{black}{51.32^0},
\end{align}
this completes the derivation of OC.

\noindent

\section{Proposed Beamforming Methods}	
\noindent
In order to reduce the interference from multi-stream and multi-user transmission, in this section, the three low-complexity closed-form beamformers, called TLL-MMSE, NSP-MTP-MRP and SO-MMSE are proposed.

\subsection{Proposed TLL-MMSE in \textcolor{black}{rank-deficient} channel}
\begin{figure}
	\centering
	\includegraphics[width=\linewidth]{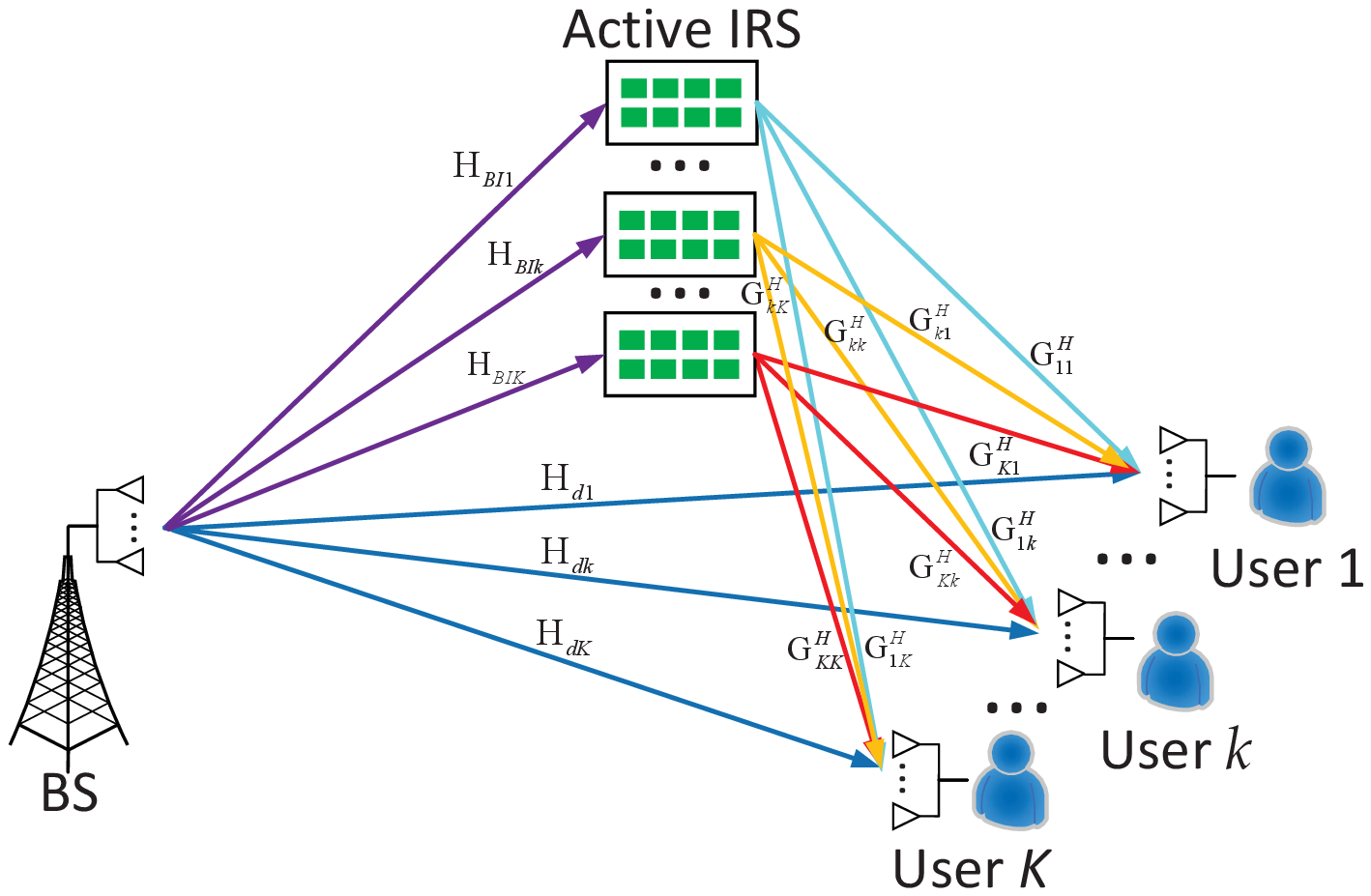}
	\caption{System model of splitting a large-scale active IRS into K smaller IRSs}
	\label{fig:DoFRy}
\end{figure}

\noindent	
In this section, assuming there is multi-stream signal transmission and $K$ users in \textcolor{black}{rank-deficient} channel. In Fig.~\ref{fig:DoFRy} the large-scale active IRS is split into $K$ smaller IRSs. Hence each IRS just has $N_k=N/K$ reflective elements. The $k$-th IRS serves the $k$-th user. At this part, a two-layer leakage plus minimum meansquare error (TLL-MMSE) is proposed. 


Then the receive signal at user k can be rewritten as
\begin{align}
	\mathbf{y}_k &= \mathbf{U}_k^H\mathbf{H}_{dk}\mathbf{V}_k\mathbf{s}_k+\mathbf{U}_k^H\sum_{j=1}^{K} \mathbf{G}_{jk}^H\mathbf{\Theta}_j\mathbf{H}_{BIj}\mathbf{V}_k\mathbf{s}_k\nonumber\\
	&+\mathbf{U}_k^H\sum_{j=1}^{K} \mathbf{G}_{jk}^H\mathbf{\Theta}_j\mathbf{H}_{BIj}\sum_{i=1,i\neq k}^{K}\mathbf{V}_i\mathbf{s}_i+\mathbf{U}_k^H\mathbf{z}_k\nonumber\\
	&+\mathbf{U}_k^H\mathbf{H}_{dk}\sum_{i=1,i\neq k}^{K}\mathbf{V}_i\mathbf{s}_i+\mathbf{U}_k^H\sum_{j=1}^{K}\mathbf{G}_{jk}^H\mathbf{\Theta}_j\mathbf{n}_k,
\end{align}
where $\mathbf{G}^H _{jk}\in \mathbb{C}^{Q_k \times N_k}$ denotes the  channel gain matrix from IRS $j$ to user $k$, $\mathbf{n}_k$ denotes the additive white Gaussian noise (AWGN) at $k$-th IRS , and $\mathbf{n}_k \sim  \mathcal {CN}(0,\textcolor{black}{\sigma^2_{irs,k}} \mathbf{I}_{N}) $.

\textcolor{black}{Then the corresponding rate can be given by}
\begin{align}
	\mathbf{\gamma}_k&= \log_2\left|\mathbf{I}_k+(\mathbf{A}_1+\mathbf{A}_2)(\mathbf{B}_1+\mathbf{B}_2+\mathbf{B}_3)^{-1}\right| ,
\end{align}

where 
\begin{align}
	\mathbf{A}_1&=\mathbf{U}_k^H\mathbf{H}_{dk}\mathbf{V}_k\mathbf{V}_k^H\mathbf{H}_{dk}^H\mathbf{U}_k,\nonumber\\ \mathbf{A}_2&=\sum_{j=1}^{K}\mathbf{U}_k^H\mathbf{G}_{jk}^H\mathbf{\Theta}_j\mathbf{H}_{BIj}\mathbf{V}_k\mathbf{V}_k^H\mathbf{H}_{BIj}^H\mathbf{\Theta}_j^H\mathbf{G}_{jk}\mathbf{U}_k,\nonumber\\
	\mathbf{B}_1&=\sum_{i=1,i\neq k}^{K}\mathbf{U}_k^H\mathbf{H}_{dk}\mathbf{V}_i\mathbf{V}_i^H\mathbf{H}_{dk}^H\mathbf{U}_k,\nonumber\\
	\mathbf{B}_2&=\sum_{j=1}^{K}\sum_{i=1,i\neq k}^{K}\mathbf{U}_k^H\mathbf{G}_{jk}^H\mathbf{\Theta}_j\mathbf{H}_{BIj}\mathbf{V}_i\mathbf{V}_i^H\mathbf{H}_{BIj}^H\mathbf{\Theta}_j^H\mathbf{G}_{jk}\mathbf{U}_k,\nonumber\\
	\mathbf{B}_3&=\sum_{j=1}^{K}\textcolor{black}{\sigma^2_{irs,k}}\mathbf{U}_k^H\mathbf{G}_{jk}^H\mathbf{\Theta}_j \mathbf{\Theta}_j^H\mathbf{G}_{jk}\mathbf{U}_k+\sigma ^2_z\mathbf{U}_k^H\mathbf{U}_k.
\end{align}

So the sum-rate of the system is figured out as follows
\begin{align}
	\mathbf{\gamma}_{sum}&=\sum_{k=1}^{K}\mathbf{\gamma}_k.
\end{align}

Let us introduce signal-to-leakage-noise ratio (SLNR). Then the  optimization problem with respect to $\mathbf{V}_k$ is presented as
\begin{align}
	&\max_{\mathbf{V}_k}~~~~	SLNR_k=\frac{A_1+A_2}{B_1+B_2+B_3}\nonumber\\
	&~~\text{s.t.}~~~~~~~~~~~~~~~~~~~\|\mathbf	{V}_k\|^2_F= P_{Tk},		
\end{align}
where $P_{Tk}$ represents the transmit power that BS allocates to $\mathbf{V}_k$, \textcolor{black}{and $P_{T}$ is the total transmit power at BS, $P_{T}=\sum_{k=1}^{K} P_{Tk}$}, then
\begin{align}
	A_1&=\|\mathbf{H}_{dk}\mathbf{V}_k\mathbf{V}_k^H\mathbf{H}_{dk}^H\|_F,\nonumber\\ A_2&=\|\mathbf{H}_{BIk}\mathbf{V}_k\mathbf{V}_k^H\mathbf{H}_{BIk}^H\|_F,\nonumber\\
	B_1&=\sum_{j=1,j\neq k}^{K}\|\mathbf{H}_{dj}\mathbf{V}_k\mathbf{V}_k^H\mathbf{H}_{dj}^H\|_F,\nonumber\\
	B_2&=\sum_{j=1,j\neq k}^{K}\|\mathbf{H}_{BIj}\mathbf{V}_k\mathbf{V}_k^H\mathbf{H}_{BIj}^H\|_F,\nonumber\\
	B_3&=\textcolor{black}{\sigma_{irs,k}^2}+\sigma ^2_z.
\end{align}

Let us define $\mathbf{A}=\mathbf{H}_{dk}^H\mathbf{H}_{dk}+\mathbf{H}_{BIk}^H\mathbf{H}_{BIk}$, $\mathbf{B}=\sum_{j=1,j\neq k}^{K}(\mathbf{H}_{dj}^H\mathbf{H}_{dj}+\mathbf{H}_{BIj}^H\mathbf{H}_{BIj})$ and $C=\textcolor{black}{\sigma^2_{irs,k}}+\sigma ^2_z$. Defining $\mathbf{V}_k=\sqrt{P_{Tk}} \hat{\mathbf{V}}_k $, then $SLNR_k$  can be reduced to the following form 
\begin{align}
	&\max_{\mathbf{V}_k}~~~~	SLNR_k=\frac{\text{tr}(\mathbf{V}_k^H\mathbf{A}\mathbf{V}_k)}{\text{tr}(\mathbf{V}_k^H\mathbf{B}\mathbf{V}_k)+C}\nonumber\\
	&~~~~~~~~~~~~~~~~~~~~~~~~~~~=\frac{\text{tr}(\hat{\mathbf{V}}^H_k \mathbf{A}\hat{\mathbf{V}}_k)}{\text{tr}(\hat{\mathbf{V}}^H_k (\mathbf{B}+\frac{C}{P_{Tk}}\mathbf{I})\hat{\mathbf{V}}_k)}\nonumber\\
	&~~\text{s.t.}~~~~~~~~~~~~~~~~~~~~~~~~\text{tr}(\hat{\mathbf{V}}_k^H\hat{\mathbf{V}}_k)=1.
\end{align}

Defining $\mathbf{D}=\mathbf{B}+\frac{C}{P_{Tk}}\mathbf{I}$, then we have
\begin{align}
	SLNR_k=\text{tr}(\hat{\mathbf{V}}^H_k\mathbf{D}^{(-1/2)H} \mathbf{A}\mathbf{D}^{(-1/2)}\hat{\mathbf{V}}_k),
\end{align}
where $\hat{\mathbf{V}}_k$ is the eigenvector corresponding to the maximum eigenvalue of $\mathbf{D}^{(-1/2)H} \mathbf{A}\mathbf{D}^{(-1/2)}$, and $\mathbf{D}^{(1/2)H}\mathbf{D}^{(1/2)}=\mathbf{D}=\mathbf{Q}^H\mathbf{F}\mathbf{Q}$, hence $\mathbf{D}^{(1/2)}=\mathbf{F}^{1/2}\mathbf{Q}$ .

Next, let us focus on the specific design of the PSM $\mathbf{\Theta}_k$ as
\begin{align}\label{T1}
	&\max_{\mathbf{\Theta}_k}	SLNR_k=\nonumber\\
	&\frac{\|\mathbf{G}_{kk}^H\mathbf{\Theta}_k\mathbf{H}_{BIk}\mathbf{V}_{k}\mathbf{s}_{k}\|^2}{\sum_{j=1,j\neq k}^{K} \|\mathbf{G}_{kj}^H\mathbf{\Theta}_k\mathbf{H}_{BIk}\mathbf{V}_{k}\mathbf{s}_{k}\|^2+\sum_{j=1}^{K}\|\mathbf{G}_{kj}^H\mathbf{\Theta}_k\|^2\sigma ^2_{irs,k}}\nonumber\\
	&~~\text{s.t.}~~~\|\mathbf{\Theta}_k\mathbf{H}_{BIk}\mathbf{V}_{k}\mathbf{s}_{k}\|^2+\|\mathbf{\Theta}_k\|^2_F\textcolor{black}{\sigma ^2_{irs,k}}\le P_{Ik},	
\end{align}
where $P_{Ik}$ represents the power at the $k$-th small IRS, and $P_{I}$ is the total power at IRS, $P_{I}=\sum_{k=1}^{K} P_{Ik}$.
Let us define $\mathbf{\Theta}_k=\text{diag}(\boldsymbol{\theta}_k)$, hence  (\ref{T1}) can be rewritten  as
\begin{align}\label{T33}
	&\max_{\mathbf{\Theta}_k}	SLNR_k =\nonumber\\
	&\frac{\boldsymbol{\theta}_k^H\textcolor{black}{\mathbf{A}_{\theta}}^H\mathbf{G}_{kk}\mathbf{G}_{kk}^H\textcolor{black}{\mathbf{A}_{\theta}}\boldsymbol{\theta}_k}{\boldsymbol{\theta}_k^H(\sum_{j=1,j\neq k}^{K} \textcolor{black}{\mathbf{A}_{\theta}}^H\mathbf{G}_{kj}\mathbf{G}_{kj}^H\textcolor{black}{\textcolor{black}{\mathbf{A}_{\theta}}}+\sum_{j=1}^{K}\textcolor{black}{\sigma ^2_{irs,k}}\mathbf{G}_{kj}\mathbf{G}_{kj}^H)\boldsymbol{\theta}_k}\nonumber\\
	&~~\text{s.t.}~~~\|\mathbf{\Theta}_k\mathbf{H}_{BIk}\mathbf{V}_{k}\mathbf{s}_{k}\|^2+\|\mathbf{\Theta}_k\|^2_F\textcolor{black}{\sigma_{irs,k} ^2}\le P_{Ik},	
\end{align}
where $\textcolor{black}{\textcolor{black}{\mathbf{A}_{\theta}}}=\text{diag}(\mathbf{H}_{BIk}\mathbf{V}_{k}\mathbf{s}_{k})$. Assuming $P_{Ik}=\|\mathbf{\Theta}_k\mathbf{H}_{BIk}\mathbf{V}_{k}\mathbf{s}_{k}\|^2+\|\mathbf{\Theta}_k\|^2_F\textcolor{black}{\sigma_{irs,k} ^2}$, and $\boldsymbol{\theta}_k=\hat{\boldsymbol{\theta}}_k \hat{\mu}_k$, where $\hat{\mu}_k=\|\boldsymbol{\theta}_k\|$, $\hat{\boldsymbol{\theta}}_k^H\hat{\boldsymbol{\theta}}_k=1$. The (\ref{T33}) can be rewritten as
\begin{align}\label{(T444)}
	&\max_{\hat{\boldsymbol{\theta}}_k}	SLNR_k
	=\nonumber\\&\frac{\hat{\boldsymbol{\theta}}_k^H\textcolor{black}{\textcolor{black}{\mathbf{A}_{\theta}}}^H\mathbf{G}_{kk}\mathbf{G}_{kk}^H\textcolor{black}{\textcolor{black}{\mathbf{A}_{\theta}}}\hat{\boldsymbol{\theta}}_k}{\hat{\boldsymbol{\theta}}_k^H(\sum_{j=1,j\neq k}^{K} \textcolor{black}{\textcolor{black}{\mathbf{A}_{\theta}}}^H\mathbf{G}_{kj}\mathbf{G}_{kj}^H\textcolor{black}{\textcolor{black}{\mathbf{A}_{\theta}}}+\sum_{j=1}^{K}\sigma ^2_{irs,k}\mathbf{G}_{kj}\mathbf{G}_{kj}^H)\hat{\boldsymbol{\theta}}_k},	
\end{align}
where $\hat{\boldsymbol{\theta}}_k$ is the eigenvector corresponding to the maximum eigenvalue of $\hat{\mathbf{D}}_k$, where
\begin{align}
	\hat{\mathbf{D}}_k=\frac{\textcolor{black}{\textcolor{black}{\mathbf{A}_{\theta}}}^H\mathbf{G}_{kk}\mathbf{G}_{kk}^H\textcolor{black}{\textcolor{black}{\mathbf{A}_{\theta}}}}{\sum_{j=1,j\neq k}^{K} \textcolor{black}{\textcolor{black}{\mathbf{A}_{\theta}}}^H\mathbf{G}_{kj}\mathbf{G}_{kj}^H\textcolor{black}{\textcolor{black}{\mathbf{A}_{\theta}}}+\sum_{j=1}^{K}\sigma ^2_{irs,k}\mathbf{G}_{kj}\mathbf{G}_{kj}^H}.	
\end{align}

And
\begin{align}
	\hat{\mu}&=\sqrt{\frac{P_{Ik}}{\hat{\boldsymbol{\theta}}_k^H(\textcolor{black}{\textcolor{black}{\mathbf{A}_{\theta}}}^H\textcolor{black}{\textcolor{black}{\mathbf{A}_{\theta}}}+\textcolor{black}{\sigma ^2_{irs,k}}\mathbf{I}_k)\hat{\boldsymbol{\theta}}_k}  },
\end{align}
then we can get $\mathbf{\Theta}_{k}$.

Next, let us fix $\mathbf{\Theta}_k$ and $\mathbf{V}_{k}$, in accordance with MMSE theorem
\begin{align}
	f(\mathbf{U}_{k})&=\mathbb{E} [(\mathbf	{y}_{k}-\mathbf	{s}_k)(\mathbf	{y}_{k}-\mathbf	{s}_k)^*]\nonumber\\
	&=\mathbb{E} [\mathbf	{y}_{k}\mathbf	{	y}_{k}^*]-\mathbb{E} [\mathbf	{	y}_{k}\mathbf	{s}_k^*]-\mathbb{E} [\mathbf	{s}_k\mathbf	{	y}_{k}^*]+1,
\end{align}
where 
\begin{align}
	&~~~~~~\mathbb{E} [\mathbf	{y}_{k}\mathbf	{	y}_{k}^*]\nonumber\\
	&=\mathbf{U}_{k}^H\mathbf{H}_{dk}\mathbf{V}_k\mathbf{V}_k^H\mathbf{H}_{dk}^H\mathbf{U}_{dk}\nonumber\\
	&+\sum_{j=1}^{K}\mathbf{U}_k^H\mathbf{G}_{jk}^H\mathbf{\Theta}_j\mathbf{H}_{BIj}\mathbf{V}_k\mathbf{V}_k^H\mathbf{H}_{BIj}^H\mathbf{\Theta}_j^H\mathbf{G}_{jk}\mathbf{U}_k\nonumber\\
	&+\sum_{i=1,i\neq k}^{K}\mathbf{U}_k^H\mathbf{H}_{dk}\mathbf{V}_i\mathbf{V}_i^H\mathbf{H}_{dk}^H\mathbf{U}_k\nonumber\\
	&+\sum_{j=1}^{K}\sum_{i=1,i\neq k}^{K}\mathbf{U}_k^H\mathbf{G}_{jk}^H\mathbf{\Theta}_j\mathbf{H}_{BIj}\mathbf{V}_i\mathbf{V}_i^H\mathbf{H}_{BIj}^H\mathbf{\Theta}_j^H\mathbf{G}_{jk}\mathbf{U}_k\nonumber\\
	&+\sum_{j=1}^{K}\textcolor{black}{\sigma^2_{irs,k}}\mathbf{U}_k^H\mathbf{G}_{jk}^H\mathbf{\Theta}_j \mathbf{\Theta}_j^H\mathbf{G}_{jk}\mathbf{U}_k+\sigma ^2_z\mathbf{U}_k^H\mathbf{U}_k,
\end{align}
and
\begin{align}
	&\mathbb{E} [\mathbf	{	y}_{k}\mathbf	{s}_{k}^*]\nonumber\\=&\mathbf{U}_{k}^H\mathbf{H}_{dk}\mathbf{V}_k+\sum_{j=1}^{K}\mathbf{U}_k^H\mathbf{G}_{jk}^H\mathbf{\Theta}_j\mathbf{H}_{BIj}\mathbf{V}_k,
\end{align}
\begin{align}
	&\mathbb{E} [\mathbf	{s}_k\mathbf	{	y}_{k}^*]\nonumber\\=&\mathbf{V}_k^H\mathbf{H}_{dk}^H\mathbf{U}_{k}+\sum_{j=1}^{K}\mathbf{V}_k^H\mathbf{H}_{BIj}^H\mathbf{\Theta}_j^H\mathbf{G}_{jk}\mathbf{U}_k.
\end{align}

Taking the derivative of function $f(\mathbf{U}_{k})$ with respect to $\mathbf{U}_{k}$ yields
\begin{align}
	\frac{ \partial f(\mathbf{U}_{k}) }{ \partial \mathbf{U}_{k} } &=2(\mathbf{H}_{dk}\mathbf{V}_k\mathbf{V}_k^H\mathbf{H}_{dk}^H\nonumber\\
	&+\sum_{j=1}^{K}\mathbf{G}_{jk}^H\mathbf{\Theta}_j\mathbf{H}_{BIj}\mathbf{V}_k\mathbf{V}_k^H\mathbf{H}_{BIj}^H\mathbf{\Theta}_j^H\mathbf{G}_{jk}\nonumber\\
	&+\sum_{i=1,i\neq k}^{K}\mathbf{H}_{dk}\mathbf{V}_i\mathbf{V}_i^H\mathbf{H}_{dk}^H\nonumber\\
	&+\sum_{j=1}^{K}\sum_{i=1,i\neq k}^{K}\mathbf{G}_{jk}^H\mathbf{\Theta}_j\mathbf{H}_{BIj}\mathbf{V}_i\mathbf{V}_i^H\mathbf{H}_{BIj}^H\mathbf{\Theta}_j^H\mathbf{G}_{jk}\nonumber\\
	&+\sum_{j=1}^{K}\textcolor{black}{\sigma^2_{irs,k}}\mathbf{G}_{jk}^H\mathbf{\Theta}_j \mathbf{\Theta}_j^H\mathbf{G}_{jk}+\sigma ^2_z\mathbf{I}_k
	)\mathbf{U}_k\nonumber\\
	&-2(\mathbf{H}_{dk}\mathbf{V}_k+\sum_{j=1}^{K}\mathbf{G}_{jk}^H\mathbf{\Theta}_j\mathbf{H}_{BIj}\mathbf{V}_k)\nonumber\\
	&=\mathbf{0},
\end{align}
which directly gives 
\begin{align}
	\mathbf{U}_{k}&=(\mathbf{H}_{dk}\mathbf{V}_k\mathbf{V}_k^H\mathbf{H}_{dk}^H+\sum_{i=1,i\neq k}^{K}\mathbf{H}_{dk}\mathbf{V}_i\mathbf{V}_i^H\mathbf{H}_{dk}^H\nonumber\\
	&+\sum_{j=1}^{K}\mathbf{G}_{jk}^H\mathbf{\Theta}_j\mathbf{H}_{BIj}\mathbf{V}_k\mathbf{V}_k^H\mathbf{H}_{BIj}^H\mathbf{\Theta}_j^H\mathbf{G}_{jk}\nonumber\\
	&+\sum_{j=1}^{K}\textcolor{black}{\sigma^2_{irs,k}}\mathbf{G}_{jk}^H\mathbf{\Theta}_j \mathbf{\Theta}_j^H\mathbf{G}_{jk}\nonumber\\
	&+\sum_{j=1}^{K}\sum_{i=1,i\neq k}^{K}\mathbf{G}_{jk}^H\mathbf{\Theta}_j\mathbf{H}_{BIj}\mathbf{V}_i\mathbf{V}_i^H\mathbf{H}_{BIj}^H\mathbf{\Theta}_j^H\mathbf{G}_{jk}\nonumber\\
	&+\sigma ^2_z\mathbf{I}_k)^{-1}(\mathbf{H}_{dk}\mathbf{V}_k+\sum_{j=1}^{K}\mathbf{G}_{jk}^H\mathbf{\Theta}_j\mathbf{H}_{BIj}\mathbf{V}_k).
\end{align}

\subsection{Proposed NSP-MTP-MRP in \textcolor{black}{LoS} channel}
 \noindent
 Assuming there are single-stream signal transmission and $K$ users. It's worth noting that IRS just aids user 1. Therefore for user 1, the received signal from the reflect channel is
\begin{align}
	y_0 &= \mathbf{u}_0^H \mathbf{G}_1^H\mathbf{\Theta}\mathbf{H}_{BI}\mathbf{v}_0 s_0+\mathbf{u}_0^H\mathbf{H}_{d1}\mathbf{v}_1 s_1\nonumber\\&+\mathbf{u}_0^H\mathbf{G}_1^H\mathbf{\Theta}\mathbf{n}+\mathbf{u}_0^H\textcolor{black}{\mathbf{z}_1},
\end{align}
where $\mathbf{v}_0$ and $\mathbf{u}_0^H$ are transmit and receive beamforming vectors of user $1$ from reflect channel respectively, and $\mathbf{\Theta} $ is the reflection coefficient matrix of IRS. And $\mathbf{v}_k $ and $\mathbf{u}_k^H $ are transmit and receive beamforming vectors of user $k$ from direct channel. 

The corresponding power is
\begin{align}
	\mathbb{E}(y_{0}^Hy_{0}) &=  |\mathbf{u}_{0}^H\mathbf{H}_{d1}\mathbf{v}_1s_1|^2+|\mathbf{u}_{0}^H \mathbf{G}_1^H\mathbf{\Theta}\mathbf{H}_{BI}\mathbf{v}_0s_0|^2\nonumber\\&+\textcolor{black}{\sigma^2_{irs}}\|\mathbf{u}_{0}^H\mathbf{G}_1^H\mathbf{\Theta}\|^2+\sigma ^2_z\|\mathbf{u}_{0}^H\|^2.
\end{align}

As thus, the corresponding signal to interference plus noise ratio (SINR) can be represented as 
\begin{align}
	SINR_{0}&=\frac{|\mathbf{u}_{0}^H \mathbf{G}_1^H\mathbf{\Theta}\mathbf{H}_{BI}\mathbf{v}_0s_0|^2}{|\mathbf{u}_{0}^H\mathbf{H}_{d1}\mathbf{v}_1 s_1|^2+A_{0}},
\end{align}
where $A_{0}=\sigma ^2_z\|\mathbf{u}_{0}^H\|^2+\textcolor{black}{\sigma^2_{irs}}\|\mathbf{u}_{0}^H\mathbf{G}_1^H\mathbf{\Theta}\|^2$.

Then the rate can be obtained as
\begin{align}
	\mathbf{\gamma}_{0}&= \log_2(1+SINR_{0})\nonumber\\
	&=\log_2\left(1+\frac{|\mathbf{u}_{0}^H \mathbf{G}_1^H\mathbf{\Theta}\mathbf{H}_{BI}\mathbf{v}_0s_0|^2}{|\mathbf{u}_{0}^H\mathbf{H}_{d1}\mathbf{v}_1 s_1|^2+A_{0}} \right).
\end{align}

The received signal from the direct channel of user $k$ is shown as
\begin{align}
	y_{k} &= \mathbf{u}_{k}^H\mathbf{H}_{dk}\mathbf{v}_k s_k+\mathbf{u}_{k}^H \mathbf{G}_k^H\mathbf{\Theta}\mathbf{H}_{BI}\mathbf{v}_0s_0\nonumber\\
	&+\mathbf{u}_{k}^H\mathbf{G}_k^H\mathbf{\Theta}\mathbf{n}+\mathbf{u}_{k}^H\textcolor{black}{\mathbf{z}_k}.
\end{align}

Similarly, the corresponding rate can be obtained 
\begin{align}
\mathbf{\gamma}_{dk}=\log_2\left(1+\frac{|\mathbf{u}_{k}^H\mathbf{H}_{dk}\mathbf{v}_ks_k|^2}{|\mathbf{u}_{k}^H \mathbf{G}_k^H\mathbf{\Theta}\mathbf{H}_{BI}\mathbf{v}_0s_0|^2+A_{k}} \right) ,
\end{align}
where $A_{k}=\textcolor{black}{\sigma^2_{irs}}\|\mathbf{u}_{k}^H\mathbf{G}_k^H\mathbf{\Theta}\|^2+\sigma ^2_z\|\mathbf{u}_{k}^H\|^2$.

For reason that user 1 has two channels to transmit signals, thus the rate of user 1 is
\begin{align}
	\mathbf{\gamma}_{1}&=\mathbf{\gamma}_{0}+\mathbf{\gamma}_{d1}.
\end{align}

For user $k$, the rate can be got directly as
\begin{align}
	\mathbf{\gamma}_{k}&=\mathbf{\gamma}_{dk},
\end{align}
where $k \ne 1$. 

Finally, the sum-rate can be rewritten as
\begin{align}
	\mathbf{\gamma}_{sum}&=\sum_{k=1}^{K}\mathbf{\gamma}_k.
\end{align}

Next, let us focus on the specific design of transmit beamforming vector $\mathbf{v}_k$, receive beamforming vector $\mathbf{u}_k$ and the phase shift matrix (PSM) $\mathbf{\Theta}$.

Defining $\mathbf{T}_k$ as the complementary channel of $\mathbf{H}_{dk}$, then  it can be shown as 
\begin{align} \mathbf{T}_k=[\mathbf{H}_{BI}^H, \mathbf{H}_{d1}^H,...,\mathbf{H}_{d(k-1)}^H,\mathbf{H}_{d(k+1)}^H,...,\mathbf{H}_{dK}^H]^H,
\end{align}
and fixing $\mathbf{u}_k$ and $\mathbf{\Theta}$, then the optimization problem with respect to $\mathbf{v}_k$ is expressed as
\begin{align}\label{5}
	&\max_{\mathbf{v}_k}~~~~	\mathbf{v}_{k}^H\mathbf{H}_{dk}^H\mathbf{H}_{dk}\mathbf{v}_{k}\nonumber\\
	&~~\text{s.t.}~~~~~~~~
	\mathbf{T}_{k}\mathbf{v}_{k}=\mathbf{0},\nonumber\\
	&~~~~~~~~~~~~~~~\mathbf{v}_{k}^H\mathbf{v}_{k}=1. 		
\end{align}

Defining $\mathbf{H}_{-k} = [\mathbf{I}-	\mathbf{T}_k^H(\mathbf{T}_k\mathbf{T}_k^H)^{\dagger}\mathbf{T}_k]$, and $\mathbf{v}_{k}=\mathbf{H}_{-k} \boldsymbol{\alpha} _k$, (\ref{5}) can be converted as
\begin{align}
	&\max_{\boldsymbol{\alpha}_k}~~~~	\boldsymbol{\alpha}_k^H\mathbf{H}_{-k}^H\mathbf{H}_{dk}^H\mathbf{H}_{dk}\mathbf{H}_{-k} \alpha _k,\nonumber\\
	&~~\text{s.t.}~~~~~~~~~~~~~~\boldsymbol{\alpha}_k^H\boldsymbol{\alpha}_k=1,	
\end{align}
according to the Rayleigh-Ritz theorem, then $\boldsymbol{\alpha}_k$ is the eigenvector corresponding to the largest eigenvalue of $\mathbf{H}_{-k}^H\mathbf{H}_{dk}^H\mathbf{H}_{dk }\mathbf{H}_{-k}$.

Then $\mathbf{v}_{k}$ can be shown as
\begin{align}\label{10}
	\mathbf{v}_{k}=\frac{\mathbf{H}_{-k}\boldsymbol{\alpha}_k}{\|\mathbf{H}_{-k}\boldsymbol{\alpha}_k\|}.
\end{align}

Similarly, we have $	\mathbf{v}_{0}=\frac{\mathbf{H}_{-0}\boldsymbol{\alpha}_0}{\|\mathbf{H}_{-0}\boldsymbol{\alpha}_0\|}$, where  $\mathbf{H}_{-0} = [\mathbf{I}-	\mathbf{T}_0^H(\mathbf{T}_0\mathbf{T}_0^H)^{\dagger}\mathbf{T}_0]$, $\mathbf{T}_0=[\mathbf{H}_{d1}^H,...,\mathbf{H}_{dk}^H,...,\mathbf{H}_{dK}^H]^H$, and  $\boldsymbol{\alpha}_0$ is the eigenvector corresponding to the largest eigenvalue of $\mathbf{H}_{-0}^H\mathbf{H}_{BI}^H\mathbf{H}_{BI }\mathbf{H}_{-0}$.

Then the optimization problem with respect to $\mathbf{u}_{k}^H$ is presented as
\begin{align}\label{6}
	&\max_{\mathbf{u}_{k}^H}~~~~	\mathbf{u}_{k}^H\mathbf{H}_{dk}\mathbf{H}_{dk}^H\mathbf{u}_{k}\nonumber\\
	&~~\text{s.t.}~~~~~~
	\mathbf{u}_{k}^H\mathbf{G}_{k}^H=\mathbf{0},\nonumber\\
	&~~~~~~~~~~~~~\mathbf{u}_{k}^H\mathbf{u}_{k}=1.		
\end{align}
Defining $\mathbf{G}_{-k} = [\mathbf{I}-	\mathbf{G}_{k}^H(\mathbf{G}_{k}\mathbf{G}_{k}^H)^{\dagger}\mathbf{G}_{k}]$, and $\mathbf{u}_{k}=\mathbf{G}_{-k} \boldsymbol{\beta} _k$, (\ref{6}) can be converted as
\begin{align}
	&\max_{\boldsymbol{\beta} _k}~~~~	\boldsymbol{\beta} _k^H\mathbf{G}_{-k}^H\mathbf{H}_{dk}\mathbf{H}_{dk}^H\mathbf{G}_{-k} \boldsymbol{\beta} _k\nonumber\\
	&~~\text{s.t.}~~~~~~~~~~~~~~\boldsymbol{\beta} _k^H\boldsymbol{\beta} _k=1,	
\end{align}
according to the Rayleigh-Ritz theorem, $\boldsymbol{\beta} _k$ is the eigenvector corresponding to the largest eigenvalue of $\mathbf{G}_{-k}^H\mathbf{H}_{dk}\mathbf{H}_{dk }^H\mathbf{G}_{-k}$ and $\mathbf{u}_{k}$ can be expressed as 
\begin{align}
	\mathbf{u}_{k}=\frac{\mathbf{G}_{-k}\boldsymbol{\beta} _k}{\|\mathbf{G}_{-k}\boldsymbol{\beta} _k\|}.
\end{align}

Similarly, we have $	\mathbf{u}_{0}=\frac{\mathbf{G}_{-0}\boldsymbol{\beta} _0}{\|\mathbf{G}_{-0}\boldsymbol{\beta} _0\|}$, where  $\mathbf{G}_{-0} = [\mathbf{I}-	\mathbf{H}_{d1}^H(\mathbf{H}_{d1}\mathbf{H}_{d1}^H)^{\dagger}\mathbf{H}_{d1}]$, and  $\boldsymbol{\beta} _0$ is the eigenvector corresponding to the largest eigenvalue of $\mathbf{G}_{-0}^H\mathbf{G}_{1}^H\mathbf{G}_{1 }\mathbf{G}_{-0}$.

The useful signal power through the reflect link at user 1 is
\begin{align}
	P=\mathbf{v}_0^H\mathbf{H}_{BI}^H\mathbf{\Theta}^H\mathbf{G}_1\mathbf{u}_{0}\mathbf{u}_{0}^H\mathbf{G}_1^H\mathbf{\Theta}\mathbf{H}_{BI}\mathbf{v}_0.
\end{align}

Let us define $\mathbf{\Theta}=\text{diag}(\boldsymbol{\theta})$, and fix $\mathbf{u}_{0}$, $\mathbf{v}_{0}$. And defining $\boldsymbol{\theta}=\hat{\boldsymbol{\theta}} \hat{\mu}$, where $\hat{\mu}=\|\boldsymbol{\theta}\|$, $\hat{\boldsymbol{\theta}}^H\hat{\boldsymbol{\theta}}=1$. Based on the phase alignment (PA) theorem, the corresponding power can be rewritten as
\begin{align}\label{15}
	P=\boldsymbol{\theta}^H\text{diag}(\mathbf{H}_{BI}\mathbf{v}_0)^H
	\mathbf{G}_1\mathbf{u}_0\mathbf{u}_0^H \mathbf{G}_1^H\text{diag}(\mathbf{H}_{BI}\mathbf{v}_0)\boldsymbol{\theta}\nonumber\\
	=\mathbf{\hat{\mu}}^2\hat{\boldsymbol{\theta}} ^H\text{diag}(\mathbf{H}_{BI}\mathbf{v}_0)^H
	\mathbf{G}_1\mathbf{u}_0\mathbf{u}_0^H \mathbf{G}_1^H\text{diag}(\mathbf{H}_{BI}\mathbf{v}_0)\hat{\boldsymbol{\theta}}.
\end{align}

Then $\boldsymbol{\theta}$ can be obtained by
\begin{align}
	\boldsymbol{\theta}&=\hat{\boldsymbol{\theta}} \hat{\mu}=\frac{\mathbf{u}_0^H \mathbf{G}_1^H\text{diag}(\mathbf{H}_{BI}\mathbf{v}_0)}{\|\mathbf{u}_0^H \mathbf{G}_1^H\text{diag}(\mathbf{H}_{BI}\mathbf{v}_0)\|}\hat{\mu}.
\end{align}

The power reflected by the IRS is
\begin{align}
	P_{I}&=\hat{\mu}^2(\|\text{diag}(\mathbf{H}_{BI}\mathbf{v}_0)\hat{\boldsymbol{\theta}}^H\|^2+\textcolor{black}{\sigma^2_{irs}}),
\end{align}
where $P_{I}$ denotes the power budget at IRS, and
\begin{align}
	\hat{\mu}&=\sqrt{\frac{P_{I}}{\|\text{diag}(\mathbf{H}_{BI}\mathbf{v}_0)\hat{\boldsymbol{\theta}}^H\|^2+\textcolor{black}{\sigma^2_{irs}}}  } .
\end{align}

\subsection{Proposed SO-MMSE in \textcolor{black}{LoS} channel} 
\noindent
In this section, there still are single-stream signal transmission and $K$ users. And IRS just aids user 1 in LoS channel. At this part, schmidt orthogonalization plus minimum mean square error (SO-MMSE) is proposed.

According to schmidt orthogonalization (SO) principle, let us define $\mathbf{v}_{0}=\mathbf{h}(\theta _{Br})$. 

Based on (\ref{10}), we have $\mathbf{v}_{1}=\frac{\mathbf{H}_{-1}\boldsymbol{\alpha}_1}{\|\mathbf{H}_{-1}\boldsymbol{\alpha}_1\|}$,
where  $\mathbf{H}_{-1} = [\mathbf{I}-	\mathbf{T}_1^H(\mathbf{T}_1\mathbf{T}_1^H)^{\dagger}\mathbf{T}_1]$, $\mathbf{T}_1=[\mathbf{H}_{BI}^H]^H$, and  $\boldsymbol{\alpha}_1$ is the eigenvector corresponding to the largest eigenvalue of $\mathbf{H}_{-1}^H\mathbf{H}_{d1}^H\mathbf{H}_{d1 }\mathbf{H}_{-1}$.

Similarly, $\mathbf{v}_{k}=\frac{\mathbf{H}_{-k}\boldsymbol{\alpha}_k}{\|\mathbf{H}_{-k}\boldsymbol{\alpha}_k\|}$,
where  $\mathbf{H}_{-k} = [\mathbf{I}-	\mathbf{T}_k^H(\mathbf{T}_k\mathbf{T}_k^H)^{\dagger}\mathbf{T}_k]$, $\mathbf{T}_k=[\mathbf{H}_{BI}^H,\mathbf{H}_{d1}^H,...,\mathbf{H}_{d(k-1)}^H]^H$, and  $\boldsymbol{\alpha}_k$ is the eigenvector corresponding to the largest eigenvalue of $\mathbf{H}_{-k}^H\mathbf{H}_{dk}^H\mathbf{H}_{dk }\mathbf{H}_{-k}$, where $2 \le k \le K$. 



Defining $\mathbf{\Theta}=\text{diag}(\boldsymbol{\theta})$, and fixing $\mathbf{v}_{0}$. And defining $\boldsymbol{\theta}=\hat{\boldsymbol{\theta}} \hat{\mu}$, where $\hat{\mu}=\|\boldsymbol{\theta}\|$, $\hat{\boldsymbol{\theta}}^H\hat{\boldsymbol{\theta}}=1$. 
The useful signal power through the reflect channel at user 1 can be rewritten as
\begin{align}
	P=\mathbf{\hat{\mu}}^2\hat{\boldsymbol{\theta}} ^H\text{diag}(\mathbf{H}_{BI}\mathbf{v}_0)^H
	\mathbf{G}_1\mathbf{G}_1^H\text{diag}(\mathbf{H}_{BI}\mathbf{v}_0)\hat{\boldsymbol{\theta}}.
\end{align}

According to the Rayleigh-Ritz theorem, $\hat{\boldsymbol{\theta}}$ is the eigenvector corresponding to the largest eigenvalue of $\text{diag}(\mathbf{H}_{BI}\mathbf{v}_0)^H
\mathbf{G}_1\mathbf{G}_1^H\text{diag}(\mathbf{H}_{BI}\mathbf{v}_0)$.
The power reflected by the IRS is
\begin{align}
	P_{I}&=\hat{\mu}^2(\|\text{diag}(\mathbf{H}_{BI}\mathbf{v}_0)\hat{\boldsymbol{\theta}}^H\|^2+\textcolor{black}{\sigma^2_{irs}}),
\end{align}
where $P_{I}$ denotes the power budget at IRS, and
\begin{align}
	\hat{\mu}&=\sqrt{\frac{P_{I}}{\|\text{diag}(\mathbf{H}_{BI}\mathbf{v}_0)\hat{\boldsymbol{\theta}}^H\|^2+\textcolor{black}{\sigma^2_{irs}}}  } .
\end{align}

Next, let us fix $\mathbf{\Theta}$, $\mathbf{v}_{0}$ and  $\mathbf{v}_{k}$, in accordance with MMSE theorem
\begin{align}
	f(\mathbf{u}_{k})&=\mathbb{E} [(\mathbf	{y}_{k}-\mathbf	{s}_k)(\mathbf	{y}_{k}-\mathbf	{s}_k)^*]\nonumber\\
	&=\mathbb{E} [\mathbf	{y}_{k}\mathbf	{	y}_{k}^*]-\mathbb{E} [\mathbf	{	y}_{k}\mathbf	{s}_k^*]-\mathbb{E} [\mathbf	{s}_k\mathbf	{	y}_{k}^*]+1,
\end{align}
where 
\begin{align}
	\mathbb{E} [\mathbf	{y}_{k}\mathbf	{	y}_{k}^*]&= \mathbf{u}_{k}^H\mathbf{H}_{dk}\mathbf{v}_k\mathbf{v}_k^H\mathbf{H}_{dk}^H\mathbf{u}_{dk}\nonumber\\
	&+\mathbf{u}_{k}^H \mathbf{G}_k^H\mathbf{\Theta}\mathbf{H}_{BI}\mathbf{v}_0\mathbf{v}_0^H\mathbf{H}_{BI}^H\mathbf{\Theta}^H\mathbf{G}_k\mathbf{u}_{k}\nonumber\\
	&+\textcolor{black}{\sigma^2_{irs}}\mathbf{u}_{k}^H\mathbf{G}_k^H\mathbf{\Theta}\mathbf{\Theta}^H\mathbf{G}_k\mathbf{u}_{k}+\sigma ^2_z\mathbf{u}_{k}^H\mathbf{u}_{k},
\end{align}
and
\begin{align}
	\mathbb{E} [\mathbf	{	y}_{k}\mathbf	{s}_{k}^*]&=\mathbf{u}_{k}^H\mathbf{H}_{dk}\mathbf{v}_k,
\end{align}
\begin{align}
	\mathbb{E} [\mathbf	{s}_k\mathbf	{	y}_{k}^*]=\mathbf{v}_k^H\mathbf{H}_{dk}^H\mathbf{u}_{k}.
\end{align}

Taking the derivative of function $f(\mathbf{u}_{k})$ with respect to $\mathbf{u}_{k}$ yields
\begin{align}
	\frac{ \partial f(\mathbf{u}_{k}) }{ \partial \mathbf{u}_{k} } &=2(	\textcolor{black}{\sigma^2_{irs}}\mathbf{G}_k^H\mathbf{\Theta}\mathbf{\Theta}^H\mathbf{G}_k+\sigma^2_z\mathbf{I}_k+\mathbf{H}_{dk}\mathbf{v}_k\mathbf{v}_k^H\mathbf{H}_{dk}^H	\nonumber\\
	&+\mathbf{G}_k^H\mathbf{\Theta}\mathbf{H}_{BI}\mathbf{v}_0\mathbf{v}_0^H\mathbf{H}_{BI}^H\mathbf{\Theta}^H\mathbf{G}_k)\mathbf{u}_k-2\mathbf{H}_{dk}\mathbf{v}_k 
\nonumber\\
&=\mathbf{0},
\end{align}
which directly gives 
\begin{align}
	\mathbf{u}_{k}&=(\mathbf{G}_k^H\mathbf{\Theta}\mathbf{H}_{BI}\mathbf{v}_0\mathbf{v}_0^H\mathbf{H}_{BI}^H\mathbf{\Theta}^H\mathbf{G}_k+\sigma^2_z\mathbf{I}_k+\nonumber\\
	&\mathbf{H}_{dk}\mathbf{v}_k\mathbf{v}_k^H\mathbf{H}_{dk}^H	+
	\textcolor{black}{\sigma^2_{irs}}\mathbf{G}_k^H\mathbf{\Theta}\mathbf{\Theta}^H\mathbf{G}_k)^{-1}(\mathbf{H}_{dk}\mathbf{v}_k).
\end{align}

Similarly, we have
\begin{align}
	&\mathbf{u}_{0}=(\mathbf{G}_1^H\mathbf{\Theta}\mathbf{H}_{BI}\mathbf{v}_0\mathbf{v}_0^H\mathbf{H}_{BI}^H\mathbf{\Theta}^H\mathbf{G}_1+\sigma^2_z\mathbf{I}_{0}+\nonumber\\
	&\mathbf{H}_{d1}\mathbf{v}_1\mathbf{v}_1^H\mathbf{H}_{d1}^H	+
	\textcolor{black}{\sigma^2_{irs}}\mathbf{G}_1^H\mathbf{\Theta}\mathbf{\Theta}^H\mathbf{G}_1)^{-1}(\mathbf{G}_1^H\mathbf{\Theta}\mathbf{H}_{BI}\mathbf{v}_0).
\end{align}

\section{Simulation Results}
\noindent
In the following, simulation results are presented to verify the DoF gains achieved by IRS and assessed the corresponding  sum-rate performance of the designed three methods. Below, simulations are conducted in two different situations: \textcolor{black}{LoS plus Rayleigh fading channels and LoS plus LoS channels, where LoS plus Rayleigh fading channels represent that the channel BS-user is LoS channel and the BS-IRS-user channel  is Rayleigh fading channel, and LoS plus LoS channels represent that the channel BS-user 
and the BS-IRS-user channel are LoS channels.}

\subsection{\textcolor{black}{LoS plus Rayleigh fading channels}}
\begin{figure}
	\centering
	\includegraphics[width=1\linewidth]{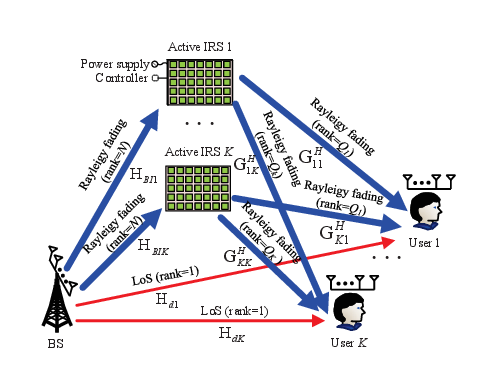}
	\caption{\textcolor{black}{System model for the scenario of LoS plus Rayleigh fading channels}}
	\label{fig:simulation-los-rayleigy}
\end{figure}
\noindent
In the section \textcolor{black}{, Fig.~\ref{fig:simulation-los-rayleigy} is the system model for the scenario of LoS plus Rayleigh fading channels. The system parameters are set as follows: $M=256$, $N=60$, $K=4$, $Q_k=4$, $P_T=40$ dBm, $P_I=30$ dBm, $\sigma_u^2=-120$ dBm, and $\sigma_{irs}^2=-90$ dBm. BS, IRS and $K$ users are respectively located at (100 m, 20 m, 0 m), (100 m, -20 m, 0 m), (100 m, 60 m, 0 m), and (100 m, -60 m, 0 m).} 

\begin{figure}
	\centering
	\includegraphics[width=1\linewidth]{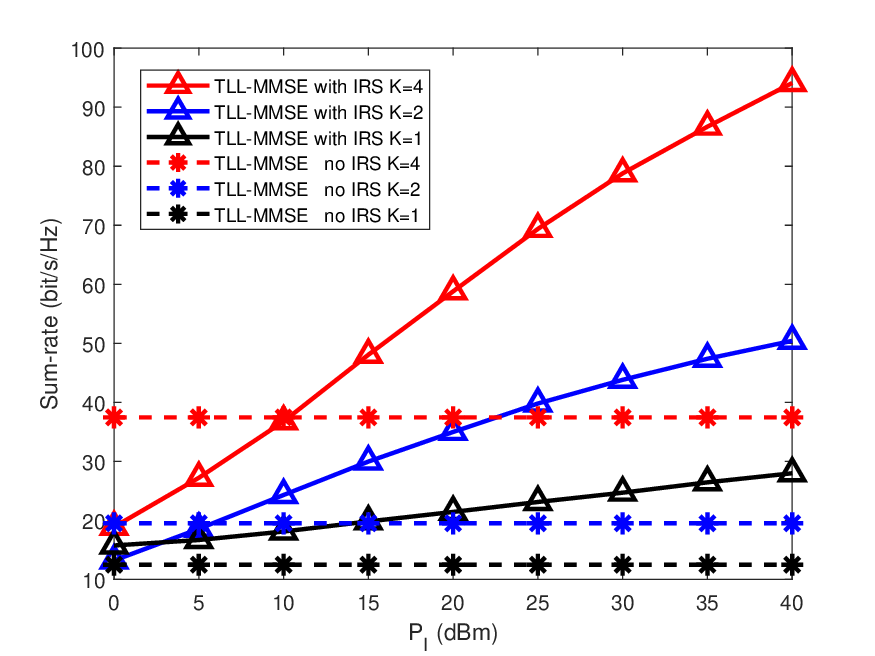}
	\caption{\textcolor{black}{Sum-rate versus IRS power $P_I$ in LoS plus Rayleigh fading channels}}
	\label{fig:pi-tll-mmse}
\end{figure}
\textcolor{black}{Fig.~\ref{fig:pi-tll-mmse} shows the curves of sum-rate versus IRS power $P_I$ for LoS plus Rayleigh fading channels under different numbers of users. Compared to no IRS, the sum-rate with IRS is higher especially in the high-power region. As $P_I$ increases from 0 dBm to 40 dBm, the sum-rate with IRS grows in general.}

\begin{figure}
	\centering
	\includegraphics[width=1\linewidth]{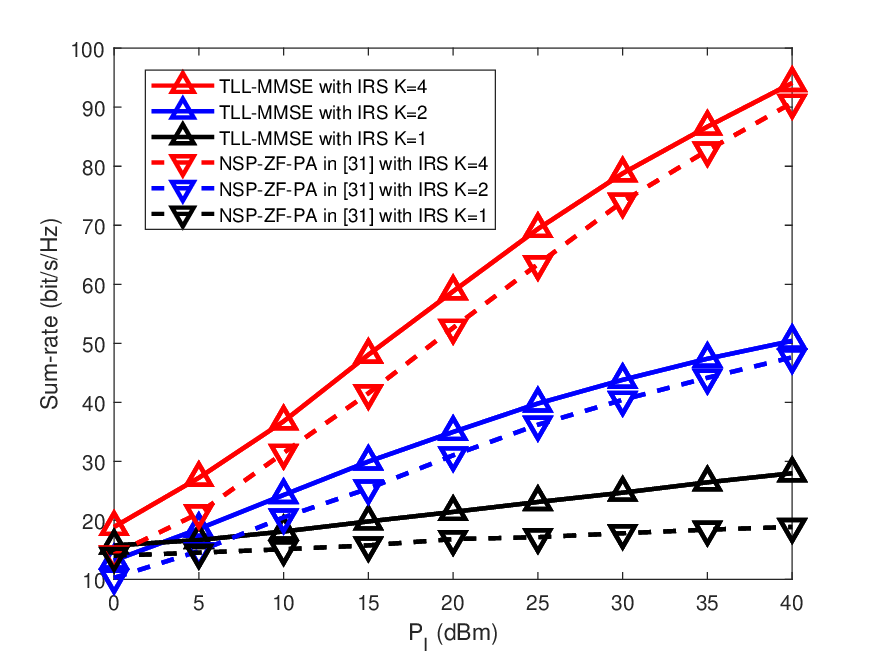}
	\caption{\textcolor{black}{Sum-rate versus IRS power $P_I$ in LoS plus Rayleigh fading channels for different methods}}
	\label{fig:nsp-zf-papi}
\end{figure}
 Fig.~\ref*{fig:nsp-zf-papi} plots the curves of sum-rate versus power $P_I$ at IRS for the proposed TLL-MMSE method and \textcolor{black}{existing NSP-ZF-PA} method in [31]. From Fig.~5, it can be seen that the sum-rate of \textcolor{black}{both methods increase} as $P_I$ grows. This indicates that \textcolor{black}{increasing the} IRS power effectively improves the communication sum-rate performance. In addition, the proposed TLL-MMSE method performs better than NSP-ZF-PA \textcolor{black}{as $P_I$ varies from 0 dBm to 40 dBm in terms of sum-rate.} In particular,  when $K=4$ and $P_I$ = 40 dBm, the sum-rate achieved by the proposed TLL-MMSE method is about 5 bit/s/Hz higher than that of NSP-ZF-PA method. It indicates that the the proposed TLL-MMSE method can effectively suppress the interference and improve the resource utilization.
	
\begin{figure}
	\centering
	\includegraphics[width=1\linewidth]{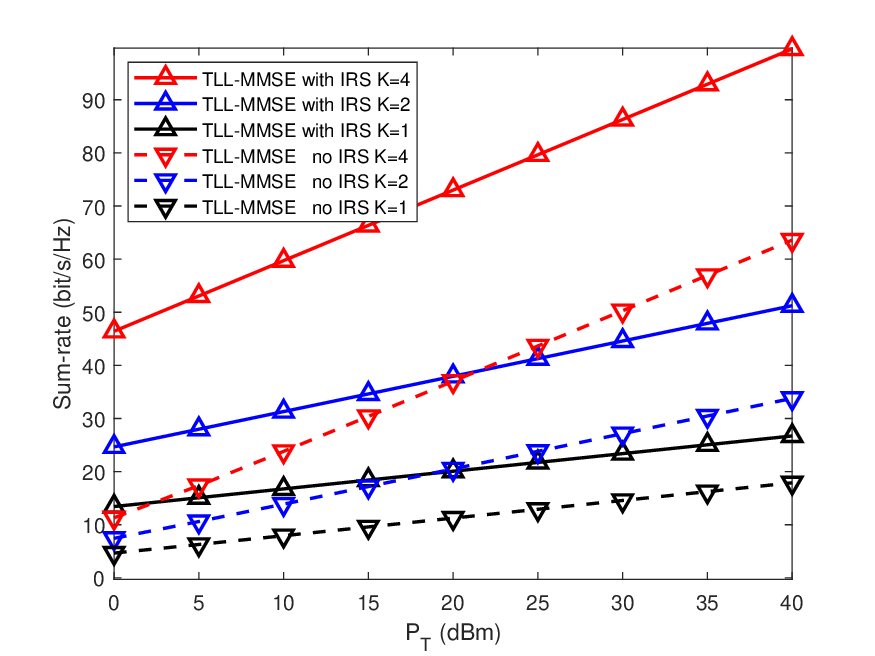}
	\caption{Sum-rate versus transmit power $P_T$ for LoS plus Rayleigh fading channels}
	\label{fig:sr-pt}
\end{figure}
Fig.~\ref{fig:sr-pt} displays the curves of sum-rate versus transmit power $P_T$ at BS for LoS plus Rayleigh fading channels with different numbers of users when $\sigma_u^2=-50$ dBm and $\sigma_{irs}^2=-70$ dBm. As $P_T$ varies from 0 dBm to 40 dBm, the sum-rate increases as $P_T$ grows in general. Compared to no IRS, the sum-rate with IRS is much higher given a fixed $P_T$. 
	
\begin{figure}
	\centering
	\includegraphics[width=1\linewidth]{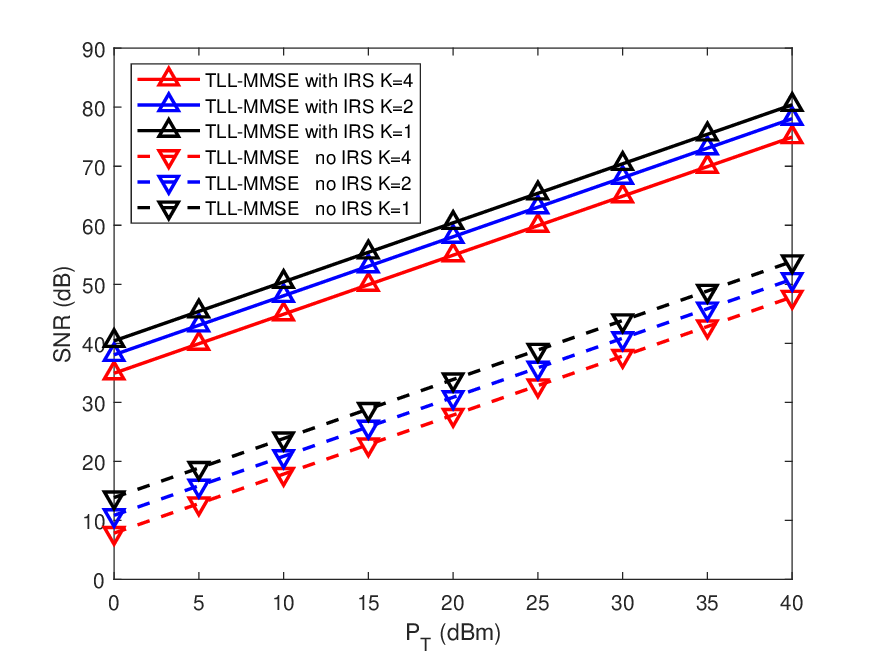}
	\caption{SNR versus transmit power $P_T$ for LoS plus Rayleigh fading channels}
	\label{fig:snr-pt}
\end{figure}

In accordance with (6), the SNRs corresponding to the different BS transmit power $P_T$ as shown in the axis-x of Fig.~\ref{fig:sr-pt} can be calculated and plotted in Fig.~\ref{fig:snr-pt}. In this figure, as the transmit power at BS $P_T$ increases, the SNR per stream grows gradually. 
Compared to no IRS, the SNR with IRS is higher. In other words, IRS may improve the receive SNR at user. 
When $P_T=$ 25 dBm, the average SNRs at users are up to 30.9 dB (no IRS ) and 63 dB (with IRS), respectively. Clearly, the two SNR values are high.
For a better explanation, the SNRs and sum-rates corresponding to the different BS transmit power when $K$=2 are listed in Table \ref{SNR-P_T1}. 

\begin{figure}
	\centering
	\includegraphics[width=1\linewidth]{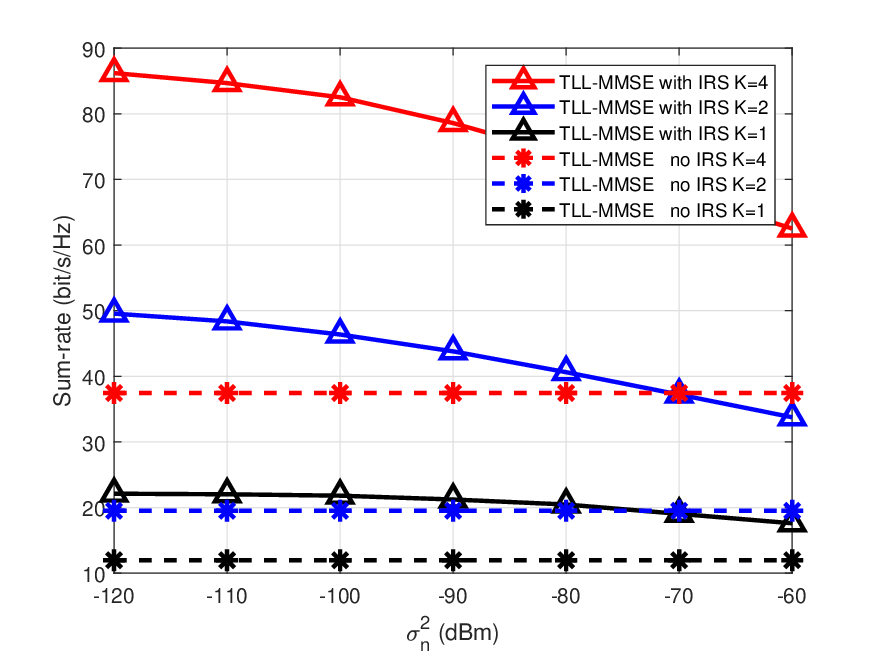}
	\caption{Sum-rate versus $\sigma_{irs}^2$ at IRS in LoS plus Rayleigh fading channels}
	\label{fig:sigma-tll-mmse}
\end{figure}

\begin{table}[!ht]
	\centering
	\caption{Sum-rate and SNR versus transmit power $P_T$ in LoS plus Rayleigh fading channels when $K=2$}
	\vspace{7pt}
	\resizebox{1\columnwidth}{!}{  
	\begin{tabular}{|c|c|c|c|c|c|c|c|c|c|c|}
		\hline
		~ & P\_T (dBm) & 0 & 5 & 10 & 15 & 20 & 25 & 30 & 35 & 40 \\ \hline
		no IRS & SNR (dB) & 10.9 & 15.9 & 20.9 & 25.9 & 30.9 & 35.9 & 40.9 & 45.9 & 50.9 \\ \hline
		no IRS & sum-rate (bit/s/Hz) & 7.4 & 10.6 & 13.9 & 17.2 & 20.5 & 23.8 & 27.1 & 30.5 & 33.8 \\ \hline
		with IRS & SNR (dB) & 38.0 & 43.0 & 48.0 & 53.0 & 58.0 & 63.0 & 68.0 & 73.0 & 78.0 \\ \hline
		with IRS & sum-rate (bit/s/Hz) & 24.7 & 28.0 & 31.3 & 34.6 & 38.0 & 41.3 & 44.6 & 48.0 & 51.2 \\ \hline
	\end{tabular}}
	\label{SNR-P_T1}
\end{table}

\textcolor{black}{Fig.~\ref{fig:sigma-tll-mmse} presents that the curves of sum-rate versus $\sigma_{irs}^2$ at IRS for the proposed TLL-MMSE method. From Fig.~\ref{fig:sigma-tll-mmse}, it is \textcolor{black}{seen} that with the aid of IRS, the sum-rate of the multi-user MIMO network decreases gradually as $\sigma_{irs}^2$ grows from -120 dBm to -60 dBm. When $\sigma_{irs}^2$ is small, the IRS-assisted multi-user MIMO network can attain a higher sum-rate compared to the network without IRS. In the high $\sigma_{irs}^2$ region, the role of IRS gradually weakens, which make the sum-rate gap between the multi-user MIMO network with IRS and without IRS gradually decrease, which means that $\sigma_{irs}^2$ at active IRS has a dominant influence on the sum-rate performance. From Table \ref{tab:irs_comparison}, it is also easy to know that the sum-rate of the IRS-assisted multi-user MIMO network is \textcolor{black}{about twice that} of the network without IRS when $\sigma_n^2 \le -90$ dBm. When $K$ = 2 and $\sigma_n^2 = -120$ dBm,  the sum-rate of the IRS-assisted multi-user MIMO network even \textcolor{black}{reaches up to 2.54 times that of} the network without IRS.}

\begin{table}
		\caption{\textcolor{black}{Sum-rate Comparison with and without IRS for $K=4$ and $K=2$ when $\sigma_n^2 \leq -90$ dBm}}
		\vspace{5pt}
	\centering
	\resizebox{\columnwidth}{!}{
		\begin{tabular}{|c|ccc|c|c|c|}
			\hline
			\(\sigma_n^2\) (dBm) & \multicolumn{3}{c|}{K=4} & \multicolumn{3}{c|}{K=2} \\
			\cline{2-7}
			& \(With~IRS\) & \(No~IRS\) & \(\textbf{ratio}\) & \(With~IRS\) & \(No~IRS\) & \(\textbf{ratio}\) \\
			\hline
			-120 & 49.5 & 19.5 & \textbf{2.54} & 49.5 & 19.5 & \textbf{2.54} \\
			-110 & 48.7 & 19.5 & \textbf{2.50} & 48.7 & 19.5 & \textbf{2.50} \\
			-100 & 46.4 & 19.5 & \textbf{2.38} & 46.4 & 19.5 & \textbf{2.38} \\
			-90  & 44.0 & 19.5 & \textbf{2.26} & 44.0 & 19.5 & \textbf{2.26} \\
			\hline
		\end{tabular}
	}
	\label{tab:irs_comparison}
\end{table}

\subsection{\textcolor{black}{LoS plus LoS channels}}
\noindent
\begin{figure}
	\centering
	\includegraphics[width=1\linewidth]{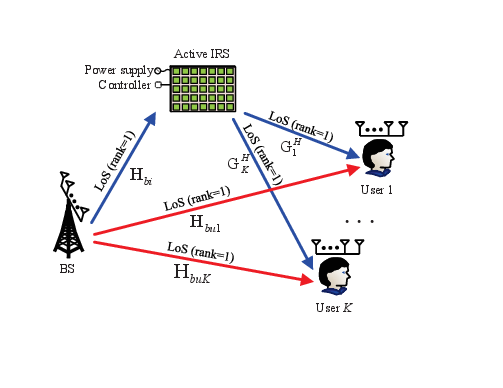}
	\caption{\textcolor{black}{System model for the scenario of LoS plus LoS channels}}
	\label{fig:simulation-loslosb}
\end{figure}
In the section all channels are assumed in LoS channel, \textcolor{black}{as shown in Fig.~\ref{fig:simulation-loslosb}. Such a scenario is a typical rank-deficient channel with each channel being extremely low-rank, i.e. rank-one.} System parameters for this part are set as follows: $\sigma^2_{irs}=\sigma_z ^2=-40$ dBm, $P_T=30$ dBm,$P_I=10$ dBm, where $P_T$ presents total transmit power from BS. $M=16$, $Q=2$.  The positions of BS and IRS are (0m, 0m, 10m) and (80m, 20m, 20m), respectively. The positions of four users are (100m, 0m, 0m), (100m, 10m, 0m), (90m, -20m, 0m) and (110m, -40m, 0m), respectively.

\begin{figure}
	\centering
	\includegraphics[width=\linewidth]{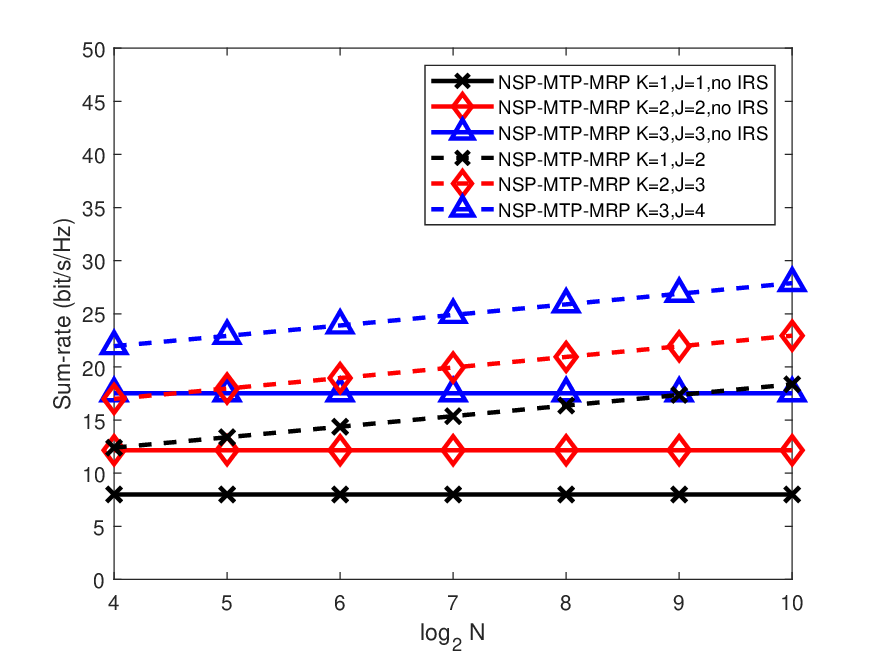}
	\caption{Sum-rate versus the number of IRS elements $N$ in \textcolor{black}{LoS plus LoS channels}}
	\label{fig:SM-N-LoS}
\end{figure}

Fig.~\ref{fig:SM-N-LoS} illustrates the sum-rate versus the number of IRS elements $N$  for the method NSP-MTP-MRP proposed by us. From this figure, it is seen that the sum-rate of the proposed NSP-MTP-MRP with IRS  has made a substantial rate enhancement over no IRS due to the increase in DoF. The former may transmit one more bit stream compared to the latter.  In particular, for the case of $K=1$ and $N=1024$, i.e., single user, the sum-rate aided  by IRS can reach up to 2.4 times that of no IRS because the number of bit streams increases from one to two in LoS channel with the help of IRS.

\begin{figure}[h]
	\centering
	\includegraphics[width=\linewidth]{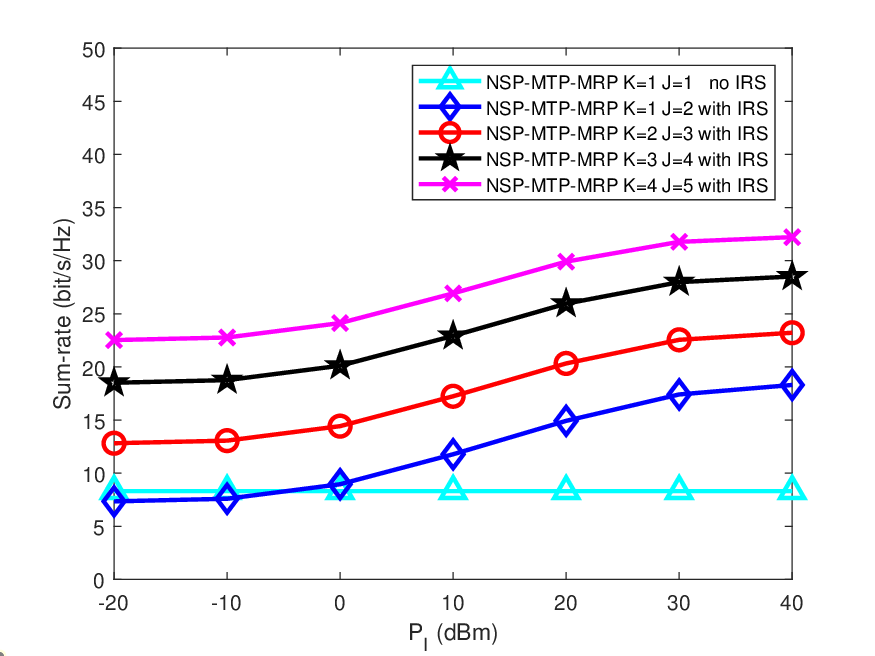}
	\caption{Sum-rate versus power $P_I$ at IRS in \textcolor{black}{LoS plus LoS channels}}
	\label{fig:SM-P_I-LoS}
\end{figure}

Fig.~\ref{fig:SM-P_I-LoS} plots the curves of the sum-rate versus power $P_I$ at IRS, where $N=16$. As $P_I$ varies from -20 dBm to 40 dBm, the sum-rate increases as $P_I$  grows in general. However, at the extremely $P_I$ region, for example, the reflective power $P_I\leq$ -10 dBm at IRS, increasing the value of $P_I$ has no impact on rate performance.  Hence, increasing $P_I$ can improve the
sum-rate.

\begin{figure}[h]
	\centering
	\includegraphics[width=\linewidth]{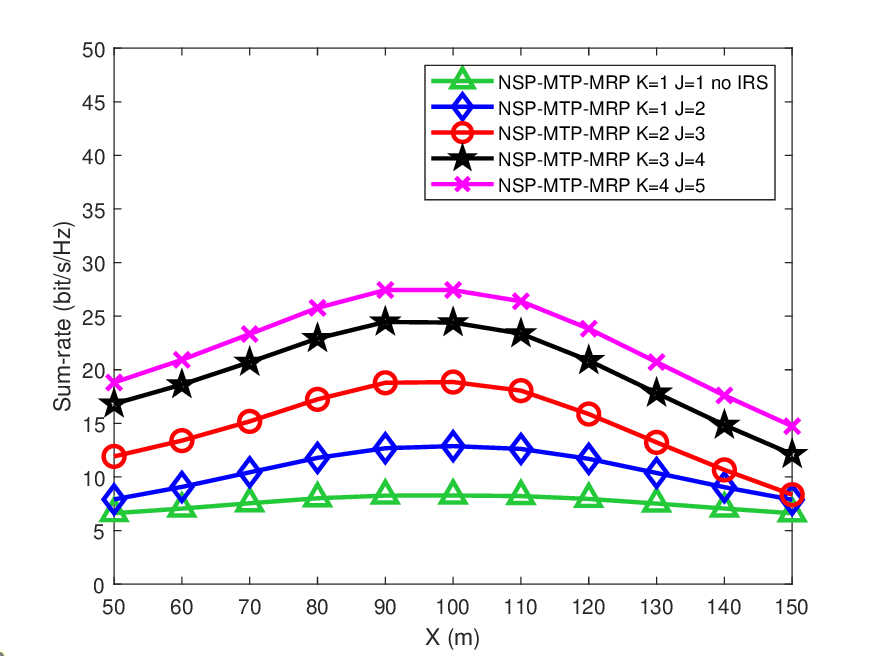}
	\caption{Sum-rate versus the x-axis value of position of IRS  in \textcolor{black}{LoS plus LoS channels}}
	\label{fig:SM-X-LoS}
\end{figure}

Fig.~\ref{fig:SM-X-LoS} illustrates the sum-rate versus the x-axis value of position of IRS $X$ when $N=16$. Its x-axis value changes  from 50m to 150m by fixing the  remaining axes  of IRS. The sum-rate function of $X$ is a concave function over the x-axis range  with the help of IRS. Obviously, there is at least one optimal position of IRS to achieve the maximum DoF. It is also easy to know that the conclusion applies to both low-SNR and high-SNR conditions.


\begin{figure}
	\centering
	\includegraphics[width=1\linewidth]{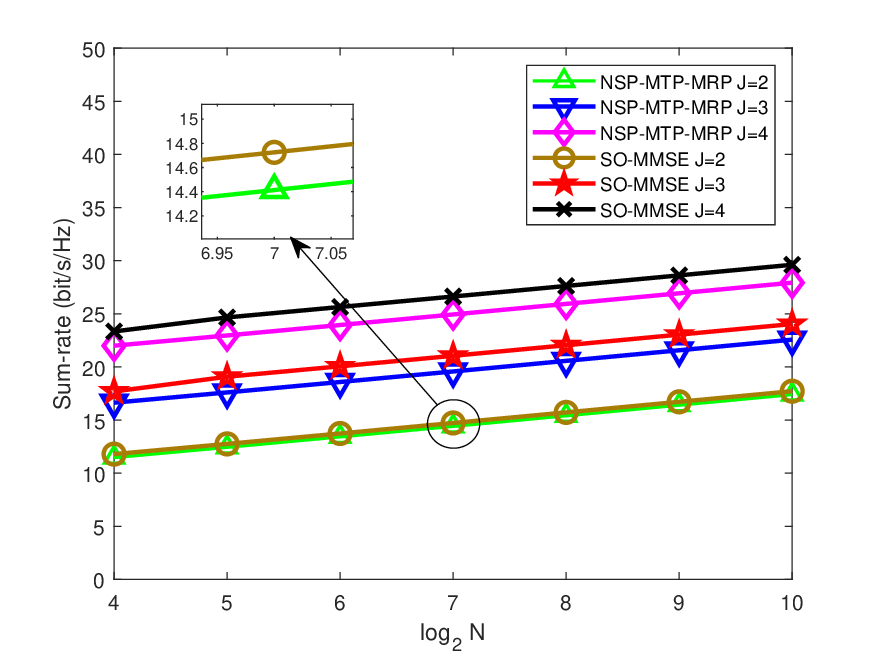}
	\caption{Sum-rate versus the number of IRS elements $N$ for two different methods in \textcolor{black}{LoS plus LoS channels}}
	\label{fig:SM-N-Ray}
\end{figure}

Fig.~\ref{fig:SM-N-Ray} makes a comparison between the sum-rates versus the number of IRS elements $N$ of the two methods, i.e., SO-MMSE and NSP-MTP-MRP. From this figure, it is seen that the SO-MMSE outperforms NSP-MTP-MRP as SO-MMSE in terms of  sum-rate. More importantly, as the number of users increases, the sum-rate gain of the former over the latter grows gradually.


 \section{Conclusions} 
  \noindent
 In this paper, the achievable optimal DoF, and obtained sum-rate gains of an active IRS-aided MU-MIMO wireless communication network are investigated in LoS and \textcolor{black}{low-rank}  channels, respectively. Specifically, we have derived the DoF upper bounds of IRS-aided single-user MIMO by using matrix inequalities and also extend the corresponding results to multi-user scenarios. \textcolor{black}{In the low-rank channel, introducing IRS will make a dramatic DoF enhancement, i.e., doubling the upper bound of DoF.} To examine the rate gains achieved by enhanced DoF, subsequently, three beamforming methods that NSP-MTP-MRP, SO-MMSE and TLL-MMSE are designed to achieve the maximum DoF. Finally, from simulation results, in \textcolor{black}{LoS plus Rayleigh fading channels}, the devised TLL-MMSE aided by IRS may achieve a \textcolor{black}{2.54 times} sum-rate over no IRS, even in the small-scale or medium-scale scenarios, for example $N$ = 60. This is mainly due to the doubled increment in DoF induced by IRS. In \textcolor{black}{LoS plus LoS channels},  the rate of the combined method NSP-MTP-MRP is more than twice that of no IRS for the SU-MIMO scenario in \textcolor{black}{LoS plus LoS channels} when the size of IRS goes to large-scale. In summary, the immanent DoF enhancement ability of IRS in low-rank channel makes it become one of the most promising techniques for the future generation wireless communication networks like 6G, particularly a kind of low-rank-channel scenarios such as satellite communication, space communications, UAV communications, and marine communications. 
 
 \vskip 2mm
 \zihao{5}
 \noindent
 \textbf{Acknowledgment}
 \vskip 2mm
 
 \zihao{5--}
 \noindent

\renewcommand\refname{\zihao{5}\textbf{References}}
This work was supported in part by the National Natural Science Foundation of China under Grant U22A2002, and  by the Hainan Province Science and Technology Special Fund under Grant ZDYF2024GXJS292; in part by the Scientific Research Fund Project of Hainan University under Grant KYQD(ZR)-21008; in part by the Collaborative Innovation Center of Information Technology, Hainan University, under Grant XTCX2022XXC07; in part by the National Key Research and Development Program of China under Grant 2018YFB180110.

  \end{document}